\documentclass[reprint,showpacs,superscriptaddress,amsmath,amssymb,aps,twocolumn,floatfix]{revtex4-1}

\usepackage{graphicx}
\usepackage{color}
\usepackage{dcolumn}
\usepackage{bm}
\usepackage{amsmath}
\usepackage{amsfonts}
\usepackage{amssymb}
\usepackage{tabularx}
\usepackage{xspace}
\usepackage{ulem}
\usepackage[caption=false]{subfig}
\usepackage{mathtools}
\usepackage[utf8]{inputenc}
\usepackage{xparse}

\newcommand\ii{\mathrm{i}}

\newcommand\ee{\mathrm{e}}

\begin{document}

\title{Bad metal and negative compressibility transitions in a two-band Hubbard model}

\author{Raymond Fr\'esard}
\altaffiliation{Raymond.Fresard@ensicaen.fr} 
\affiliation{Normandie Universit\'e, ENSICAEN, UNICAEN, CNRS, CRISMAT, 14050 Caen, France}
\author{Kevin Steffen}
\affiliation{Center for Electronic Correlations and Magnetism, EP VI, Institute of Physics, University of Augsburg, 86135 Augsburg, Germany}
\author{Thilo Kopp}
\affiliation{Center for Electronic Correlations and Magnetism, EP VI, Institute of Physics, University of Augsburg, 86135 Augsburg, Germany}

\begin{abstract}
We analyze the paramagnetic state of a two-band Hubbard model with finite Hund's coupling close to integer filling at $n=2$ in two spacial dimensions. Previously, a Mott metal-insulator transition was established at $n=2$ with a coexistence region of a metallic and a bad metal state  in the vicinity of that integer filling. The coexistence region ends at a critical point beyond which a charge instability persists.
Here we investigate the transition into negative electronic compressibility states for an extended filling range close to $n=2$ within a slave boson setup. We analyze the separate contributions from the (fermionic) quasiparticles and the (bosonic) multiparticle incoherent background and find that the total compressibility depends on a subtle interplay between the quasiparticle  excitations and collective fields. Implementing a Blume-Emery-Griffiths model approach for the slave bosons, which mimics the bosonic fields by Ising-like pseudospins, we  suggest a feedback mechanism between these fields and the fermionic degrees of freedom.
We argue that the negative compressibility can be sustained for heterostructures of such strongly correlated planes and results in a  large capacitance of these structures. The strong density dependence of these capacitances allows to tune them through small electronic density variations. Moreover, by resistive switching from a Mott insulating state to a metallic state through short electric pulses, transitions between fairly different capacitances are within reach.
\end{abstract}

\date{\today}

\maketitle

\section{Introduction.}
\label{sec:intro}

Strongly correlated electron systems have been in the focus of research for
many decades, not the least on account of their peculiar
magnetic~\cite{Fazekas99} and unconventional superconducting
properties~\cite{Anderson87,Norman11,Scalapino12,Keimer15}. 
The  manifest characteristic of the prominent model for strongly correlated electrons, the one-band Hubbard model~\cite{Hubbard1963,Gutzwiller1963}, is the doping-driven Mott metal-insulator transition (MIT)~\cite{Brinkman1970}. It is the repulsive on-site Coulomb interaction $U$ that renders a transition into a Mott insulating state at half-filling ($n=1$).
With respect to MITs, an extension of the model to a multiband case~\cite{Kanamori63,Lu1994,Fresard1997,Lechermann2007} appears to be qualitatively similar except that insulating states are to be identified at integer filling numbers. For example, in the case of two orbitals per site insulating states can emerge at $n=1,2,3$ ---apart from the uncorrelated insulating states at $n=0$ and 4. 

However, an unsophisticated reasoning with respect to multi-band behavior must
fail on several accounts: For asymmetric two-orbital Hubbard models,
presenting systems with unequal local Coulomb interactions for distinct
orbitals or different band widths, orbital selective Mott phases are to be
expected where one band may be insulating whereas the second band is metallic
(see, for example,
Refs.~[\onlinecite{Anisimov2002,Ruegg05,Biermann2005,Liebsch2005,Koga2005}]). 
Moreover, when further coupling parameters become relevant, such as Hund's
coupling $J_{\rm H}$, various magnetic phases are
stabilized~\cite{Hotta04,Fre05n,Rac06n,Quan2018}. 
Recently, an in-gap band for the two-orbital case~\cite{Hallberg18} has been
identified, the width of which depends on $J_{\rm
  H}$~\cite{Hallberg21}. Furthermore, even for modest Coulomb interaction $U$,
Hund's coupling $J_{\rm H}$ may strongly reduce the coherence of the
underlying metallic state. This
  prominently applies to the degenerate three-band Hubbard model around one
  charge away from half-filling, in the so-called Hund metal regime \cite{Werner08, Fanfarillo15, Stadler19}.

Intriguing
is also the nature of electronic phases in the vicinity of the insulating states at 
these critical filling factors. If $U$ is on the order of its critical value or above, the
electronic state is a bad metal state with correlation-suppressed band
width. For a two-band Hubbard model with finite $J_{\rm H}$, close to $n=2$, a first order transition
was established from a moderately correlated metallic state into a bad metal, where the transition and the coexistence regime strongly depend on $J_{\rm H}/U$ and end at a critical point~\cite{Fresard2001}. In particular, the quasiparticle weight $z^2(n)$ collapses to a small value at this transition and a finite Hund's coupling controls this behavior in the two-orbital case, as correlations then depend on the local spin alignment. 

In our work we focus on the transitions into the bad metal behavior and into a
negative compressibility state in the vicinity of  $n=2$ for a symmetric
two-band Hubbard model with finite coupling $J_{\rm H}$.  Beyond the first
order transition into the bad metal regime~\cite{Fresard2001}, a continuous transition---at which
the electronic compressibility diverges indicating a charge instability---was previously
identified~\cite{Medici17}. 

A different scenario for a strongly enhanced or negative compressibility in a three-band model 
was suggested  for the insulator-metal 
transition in Sr-doped LaTiO$_3$~\cite{Liebsch2008}. There, an interorbital charge transfer may result in a negative  subband  compressibility, assuming that at least one band is close to a Mott transition.
Furthermore, we note that in the low density regime, Coulomb interactions dominate the kinetic energy and generate a negative compressibility of the electron gas~\cite{Bello81,Tanatar89}. These scenarios are not covered
by our present work.

Transitions into a state of negative electronic compressibility were  observed experimentally at interface electron gases in Si-MOSFETs and in III-V heterostructures~\cite{Eisenstein92,Kravchenko89,Shapira96}.
Moreover,  electron liquids formed at LaAlO$_3$--SrTiO$_3$ interfaces through electronic reconstruction
may allow for negative compressibility \cite{Mannhart11} as confirmed in Kelvin probe microscopy measurements~\cite{Tinkl12}.

It should be noted that a negative electronic
compressibility does not necessarily imply a thermodynamic instability---with a
possible transition into a phase separated state: the negative (inverse) 
compressibility may be compensated by positive terms which are generically given by
the ionic background or by coupling to further electronic systems, as realized in some heterostructures. 
In this case the transition into a state of negative compressibility may be continuous.
Here we do not investigate the nature of the negative compressibility state. It depends on the material and the interplay between
local and long range Coulomb interaction. Usually it is expected that the electronic system phase separates or a CDW state is formed. However, these may be exponentially damped~\cite{Schakel01} and the state stays rather homogeneous with a negative compressibility as at LaAlO$_3$--SrTiO$_3$ interfaces.

Not surprisingly, in a one-band Hubbard model the compressibility of the paramagnetic state is reduced
with respect to its free electron value and stays positive, yet strikingly
the compressibility is a non-monotonous function of $U$ for electron densities
in proximity to half-filling. The reduction is controlled by the interplay of
the effective mass and the Landau parameter $F_0^s$~\cite{Vollhardt84,Steffen16,Steffen17}. The
same is true in the vicinity of the MITs at $n=1,3$ in the two-band Hubbard
model but the case of $n=2$ is fascinatingly different. There, a finite $J_{\rm H}$
aligns the spins in the two different orbitals of a site which induces a
suppression of orbital fluctuations in the vicinity of
$n=2$ and strongly enhances the effective mass~\cite{Fresard2001,Medici17}. Nevertheless, it is remarkable that a
repulsive local Coulomb interaction  induces a negative compressibility
state. 

Here, we analyze the interplay of quasiparticle behavior, expressed by the quasiparticle
weight $z^2(n)$, and collective excitations, expressed by bosonic fields for
orbital occupations in the two-band Hubbard model. 
The feedback between these fermionic and bosonic degrees of freedom  determines
 the discontinuous and continuous phase transitions and drives the
 electronic system into a state of negative compressibility.

The slave boson technique is well adjusted to study this interplay.
In fact, negative electronic compressibility obtained by means of 
Kotliar-Ruckenstein and related slave boson calculations received considerable attention in the
context of the Hubbard model on the square lattice. In the course of considering
incommensurate spiral phases---which allow to lower the energy of the lightly doped one-band
Hubbard model with respect to the commensurate antiferromagnetic phase---
negative compressibility in a small density range close to half
band filling  was discovered~\cite{Fre91}. Motivated by the quest of thermodynamically stable
phases a Maxwell construction was suggested, which has been recently
revisited \cite{Seufert2021}. 

A well accessible response function to probe the compressibility is the
capacitance of heterostructures comprising two electrodes and dielectric
layers in between. 
An enhancement of the capacitance in two-band systems was suggested in
Ref.~\cite{Kopp09}. Besides, the capacitance of multilayers with strongly
correlated materials was investigated recently
\cite{Hale12,Freericks16,Steffen17}, either with a barrier or electrodes
consisting of strongly correlated materials. The capacitance strongly depends
on the correlation strength $U$ of the considered one-band models.
Apart from these analyses, a scheme that builds on a Wigner crystal-like strongly correlated liquid state was proposed for the low density regime to explain capacitance enhancements~\cite{Skinner10}.

In the present work we suggest a realization of a capacitance device which
comprises plates with a material that is electronically  in a regime well
described by a two-band Hubbard model close to half filling. 

 The paper is organized as follows: the two-band Hubbard model of our
 investigation is presented in Sec.~\ref{sec:model}, together with the key
 features of the extended Kotliar-Ruckenstein slave-boson technique that we
 utilize. We present our results  in Sec.~\ref{sec:resu}. These comprise the
 quasiparticle residue $z^2$ and the phase diagram close to half filling in
 Sec.~\ref{sec:QP_and_PD}, then the double occupancies as represented by slave
 boson fields in Sec.~\ref{sec:SBF}, the electronic compressibility $\kappa$
 in Sec.~\ref{sec:compressibility}, and eventually the capacitance of a device
 with strongly correlated electron systems on the electrodes in
 Sec.~\ref{sec:cap}. In Sec.~\ref{sec:BEG} the bosonic degrees of freedom are
 interpreted in terms of classical Ising-fields through a
 Blume-Emery-Griffiths (BEG) model approach, and a feedback mechanism between
 these fields and the fermionic degrees of freedom is presented. Finally,
 Sec.~\ref{sec:conclusion} presents conclusions and a short outlook. 
 
 The gauge symmetry group of the approach is unraveled in
 Appendix~\ref{app_GS}, while the saddle point equations that we solve are
 detailed in Appendix~\ref{app_SP}. The filling dependence of the chemical
 potential is given in Appendix~\ref{app_mu} and the single and triple
 occupancies are addressed in Appendix~\ref{app_st}. The parameters that enter
 the BEG-type analysis are discussed in Appendix~\ref{app_BEG_parameters}, and
 the BEG phase diagram for the chosen set of parameters  in
 Appendix~\ref{app_BEG_phase_diagram}.

\section{Model and Method}
\label{sec:model}

The microscopical model consists of a kinetic
energy term $\hat{H}_0$ and a Hubbard interaction part $\hat{H}_i$, with the complete Hamiltonian
$\hat{H}=\hat{H}_0+\hat{H}_i$. The kinetic term  reads for the two-band case
\begin{align}\label{hamiltonian_0}
\hat{H}_0&=\sum_{{\bf k},\sigma}\left(c^\dagger_{{\bf k},\eta,\sigma},c^\dagger_{{\bf k},\xi,\sigma}\right)\begin{pmatrix} \varepsilon_{\eta,{\bf k}} & r_{\bf k} \\ r_{\bf k} & \varepsilon_{\xi,{\bf k}} \end{pmatrix}\begin{pmatrix}c_{{\bf k},\eta,\sigma} \\ c_{{\bf k},\xi,\sigma} \end{pmatrix}.
\end{align}
A specific realization one may
wish to consider is provided by oxides with two bands active at the Fermi energy. 
Below, we focus on  degenerate $d_{xz}$ and $d_{yz}$
orbitals dispersing on a square lattice  in
the $x$-$y$ plane---with lattice constant $a$. In that case, a minimal tight-binding model entails
$\varepsilon_{\eta/\xi,{\bf k}} = -2 t\cos{(k_{x/y}a)}$---representing 
the hopping along the proper bond---with minimal mixing 
$r_{\bf k}=-4t'\sin{(k_xa)}\sin{(k_ya)}$---arising from the hopping along the diagonals. The operators
$c^\dagger_{{\bf k},\eta,\sigma}$ 
($c^\dagger_{{\bf k},\xi,\sigma}$) create a Bloch eigenstate with wave
vector ${\bf k}$ and spin projection $\sigma$ in band $\eta$ ($\xi$). Below we
refer to a band index $u$ that takes the values $\eta, \xi$.

The band structure is appropriate to the layered
  Sr$_2$RuO$_4$ material, that crystallizes in the Ruddlesden-Popper structure
and the degeneracy of the ${\rm t}_{\rm 2g}$ multiplet is partially lifted~\cite{Noce99}.
As $t'$ is expected to be much
  smaller than $t$ we use the representative value $t'/t = 1/25$ in our
  numerical evaluations. As for $|t'/t| \leq \frac14$
the bandwidth $W$ is given by $W=4t$, we will use from now on $W$ as the band parameter instead of $t$. Our results do not depend qualitatively on this choice of $t'/t $ but rather on the relative magnitudes of the band width, Hund's coupling $J_{\rm H}$  and on-site Coulomb interaction $U_{\rm P/A/H}$ (see below).

The two non-interacting bands
$\epsilon^{(0)}_{{\bf k},\nu}$ follow as 
\begin{equation}\label{dispersion_0}
\epsilon^{(0)}_{{\bf k},\nu}=\frac{1}{2}\left(\varepsilon_{\eta,{\bf k}} +
  \varepsilon_{\xi,{\bf k}} + \nu\sqrt{\left(\varepsilon_{\eta,{\bf k}} -
  \varepsilon_{\xi,{\bf k}}\right)^2+4\,r_{\bf k}^2}\right) 
\end{equation}
with $\nu = \pm 1$.
While for the most common dispersions on the square lattice
  the van Hove singularity is located close to---or even at---half-filling, this
  is not the case with the here chosen dispersion. Having van Hove
  singularities in the relevant doping regime would suppress the kinetic
  contribution to the inverse compressibility very effectively~\cite{Kopp09}. The
  interference of this single particle effect with the correlation driven
  impact on the compressibility, studied in Ref.~\cite{Steffen17}, is avoided here thanks to the dispersion
  Eq.~(\ref{dispersion_0}). 

For the local part of the Hamiltonian,
\begin{align}\label{hamiltonian_i}
\hat{H}_i&= U_{\rm P}\sum_{i,\sigma}\hat{n}_{i,\eta,\sigma}\hat{n}_{i,\xi,\sigma}+U_{\rm A}\sum_{i,\sigma}\hat{n}_{i,\eta,\sigma}\hat{n}_{i,\xi,-\sigma}\notag\\ &\qquad +U_{\rm H}\sum_{i,u=\eta,\xi}\hat{n}_{i,u,\uparrow}\hat{n}_{i,u,\downarrow}
\end{align}
the interactions of the electrons between different bands are taken into
account: The first term originates from the interaction of electrons in
different orbitals with parallel spins and the second term from the interaction
between electrons in different orbitals with antiparallel spins. The last
term, which also appears in the single-band Hubbard model, is due to the
on-site repulsion between two electrons in the same
band. Above, $\hat{n}_{i,u,\sigma}=c^\dagger_{i,u,\sigma}c_{i,u,\sigma}$ is the
number operator on site $i$, in band $u$ and spin projection $\sigma$. For an
ion in the  octahedral environment, assumed here, the coefficients of the
interaction are related by $U_{\rm A}=U_{\rm P}+J_{\rm H}$ and
$U_{\rm H}=U_{\rm P}+3J_{\rm H}$~\cite{Sugano70,Fresard1997,Buenemann98}. 

As argued
in Ref.~[\onlinecite{Medici17}] further contributions from Hund's coupling are of minor relevance
for the considered regime. They are not considered in this work. We comment on the reduction
of Hund's coupling to Zeeman-like spin-density correlations and the value of $J_{\rm H}$  in the conclusions, Sec.~\ref{sec:conclusion}.

We use an extended Kotliar-Ruckenstein slave-boson
technique~\cite{Kotliar1986} to treat the above defined two-band
Hamiltonian. One slave-boson field is introduced for each of the sixteen
possible atomic configurations~\cite{Fresard1997}, as well as four fermionic
fields $f_{i,\alpha}$. The physical electron annihilation operators may be
expressed in terms of auxiliary particles as:
\begin{equation}\label{Eq:forc}
c_{i,\alpha}=z_{i,\alpha}f_{i,\alpha} \,, 
\end{equation}
where
$\alpha=(u, \sigma)$ is a four-valued spin-band index and
$z_{i,\alpha}$ is a combination of bosonic operators as given in 
Ref.~\cite{Fresard1997} (see also Appendix~\ref{app_GS}). A bosonic field $e$
($\varpi$) is associated to empty 
(fourfold occupied) sites, and four bosonic fields $p_{\alpha}$ ($t_{\alpha}$)
are associated to each singly (triply) occupied sites whereby the $\alpha$-state
is  filled (empty). The six different double occupancies are tied to bosons
$d_{\alpha,\alpha'}$, with $\alpha <  \alpha'$. All auxiliary fermionic and
bosonic fields satisfy canonical commutation relations, while the physical
electron operators do so provided the following constraints are satisfied:
\begin{align}
1 &= e^{\dagger}_{i} e^{\phantom{\dagger}}_{i} 
+\sum_{\alpha}p_{i,\alpha}^{\dagger}p^{\phantom{\dagger}}_{i,\alpha}+
\sum_{\alpha< \alpha'}d_{i,\alpha\alpha'}^{\dagger}d^{\phantom{\dagger}}_{i,\alpha\alpha'}\nonumber\\
&+\sum_{\alpha}t_{i,\alpha}^{\dagger}t^{\phantom{\dagger}}_{i,\alpha}+\varpi_{i}^{\dagger}\varpi^{\phantom{\dagger}}_{i}\label{lambda}\\
f^\dagger_{i.\alpha}f^{\phantom{\dagger}}_{i,\alpha}&=p_{i,\alpha}^{\dagger}p^{\phantom{\dagger}}_{i,\alpha}+
\sum_{\alpha'< \alpha}d_{i,\alpha'\alpha}^{\dagger}d^{\phantom{\dagger}}_{i,\alpha'\alpha} +
\sum_{\alpha'> \alpha}d_{i,\alpha\alpha'}^{\dagger}d^{\phantom{\dagger}}_{i,\alpha\alpha'}
\nonumber\\
&+\sum_{\alpha'\neq\alpha}t_{i,\alpha'}^{\dagger}t^{\phantom{\dagger}}_{i,\alpha'}+\varpi_{i}^{\dagger}\varpi^{\phantom{\dagger}}_{i}\label{lambda-alpha}\
\end{align}
In an imaginary time functional integral the constraints (\ref{lambda}) and
(\ref{lambda-alpha}) are incorporated in the Lagrangian together with the
(Lagrange multiplier) constraint fields $\lambda'$ and $\lambda_{\alpha}$,
respectively. Ideally the functional integrals should be calculated
exactly. Regarding spin models this has been achieved for the Ising chain
\cite{Fre01}, but in the case of interacting electron models exact evaluations
could be performed on small clusters only, either using the Barnes
representation \cite{Kop07}, or the Kotliar and Ruckenstein representation
\cite{Kop12}. Yet, such a calculation remains challenging on lattices of higher
dimensionality, and we rather resort to the saddle-point approximation.

Below, we consider the paramagnetic saddle-point
approximation obtained after having integrated out the fermionic fields (for
formal aspects of the approach see Appendix~\ref{app_SP}). In
the paramagnetic phase one may introduce $d^2_{\rm P}$, $d^2_{\rm A}$, and
$d^2_{\rm H}$,  through the relations
$d^2_{\eta\uparrow,\xi\uparrow}=d^2_{\eta\downarrow,\xi\downarrow}\equiv
d^2_{\rm P}$,
$d^2_{\eta\uparrow,\xi\downarrow}=d^2_{\eta\downarrow,\xi\uparrow}\equiv
d^2_{\rm A}$, 
$d^2_{\eta\uparrow,\eta\downarrow}=d^2_{\xi\uparrow,\xi\downarrow}\equiv
d^2_{\rm H}$, as well as $p^2$, $t^2$, and $\lambda$ through
$p^2\equiv p_{\alpha}^2$, $t^2\equiv t_{\alpha}^2$, and $\lambda \equiv
\lambda_\alpha\forall\alpha$. In terms of them, the grand potential may be
written as:
\begin{align}
\Omega/N_{\rm L} = & 2\left(U^{\phantom{2}}_{\rm P} d^2_{\rm P} + U^{\phantom{2}}_{\rm A} d^2_{\rm A} +U^{\phantom{2}}_{\rm H} d^2_{\rm H} \right)\notag\\
& + 2\left(U^{\phantom{2}}_{\rm P} + U^{\phantom{2}}_{\rm A} +U^{\phantom{2}}_{\rm H} \right) \left(2t^2 + \varpi^2\right)\notag\\
&+ \lambda' \left( e^2 \! + 4 p^2  \! + 2 \left(d^2_{\rm P} + d^2_{\rm A} +
    d^2_{\rm H}\right) + 4t^2 \! + \!\varpi^2 \!- \! 1\right)\notag\\
& - 4\lambda\left( p^2 + d^2_{\rm P} + d^2_{\rm A} + d^2_{\rm H} + 3 t^2 +\varpi^2\right)\notag\\
&-\frac{2}{\beta}\frac{1}{N_{\rm L}}\sum_{{\bf k},\nu}
\ln{\left(1+e^{-\beta E_{{\bf k},\nu}}\right)} \label{eq:Omega}.
\end{align}
In order to disburden the notation we use $U\equiv U_{\rm P}$ below.
Outside the strong coupling regime $U>5~W$ in which the $\varpi$ boson
representing the four-fold occupancy was neglected, all bosons were retained
in our calculations performed for $n\leq 2$. Results for $n\geq 2$ are obtained using particle-hole symmetry. Here
$\beta=1/k_{\rm B}T$ incorporates the temperature $T$ and
$N_{\rm L}$ is the number of lattice sites. The dispersion for the
quasiparticles is given by 
\begin{subequations}
\begin{align}	
E_{{\bf k},\nu}&=z^2\epsilon^{(0)}_{{\bf k},\nu}-\mu_{\rm eff}  \label{eq:Ek}\\
\mu_{\rm eff}&= \mu-\lambda\,.
\end{align}
\end{subequations}

For more details about the saddle-point equations see Appendix~\ref{app_SP}.

\section{Results}
\label{sec:resu}

The transition to a negative compressibility in proximity to half filling is a
remarkable feature of the two-band Hubbard model~\cite{Medici17}, a property
which is not found for the one-band Hubbard model. A first order phase
transition to a bad metal state  close to half-filling was identified before~\cite{Fresard2001}.
There the quasiparticle residue of the charge carriers drops significantly to low
values, concomitant with a jump of the effective mass to large values. These transitions are
controlled by Hund's  coupling $J_{\rm H}$, that is, they are absent for vanishing  $J_{\rm H}$.
 We emphasize that  $J_{\rm H}$ does not scale with
$U$ but rather depends on the orbital character of the electrons. 
For our investigation of the impact of intermediate to  strong correlations
on the electronic compressibility
we consider $J_{\rm H}$ of the order of $t$, namely we fix
$J_{\rm H} = W/6$. A $J_{\rm H}/U$ dependence was discussed in Ref.~\cite{Fresard2001,Medici17}.
The full  $J_{\rm H}$ dependence will be the scope of a different work.

\begin{figure}[t]
\includegraphics[scale=0.50]{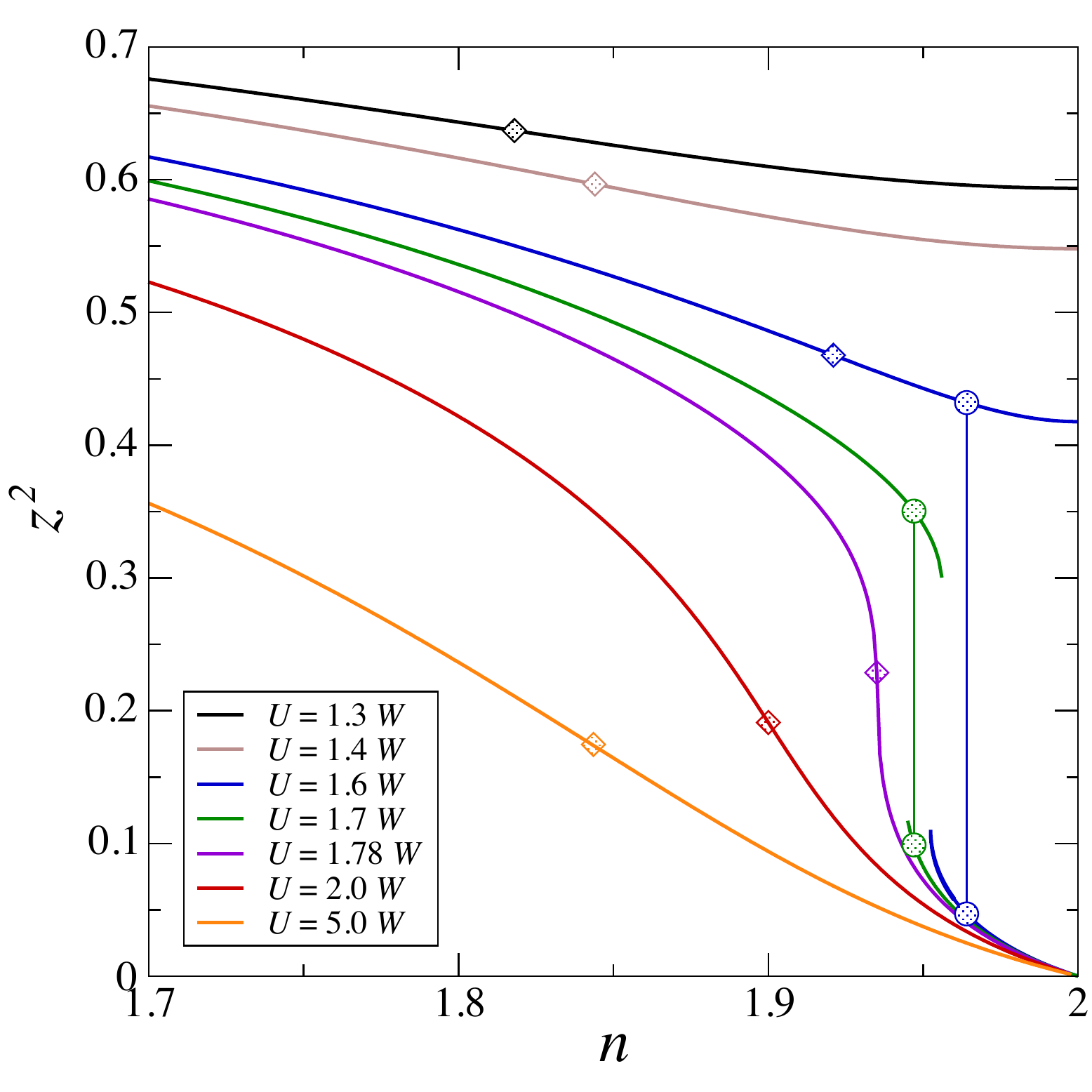}
\caption{Quasiparticle residue in dependence on filling $n$ for  $J_{\rm H} = W/6$. 
The circles and
  the vertical thin
  lines characterize the first order transitions while the diamonds mark the
  inflection points. 
  } 
\label{Fig:z}
\end{figure}

\subsection{Quasiparticle residue and phase diagram}
\label{sec:QP_and_PD}

\begin{figure}[t]
\includegraphics[scale=0.54]{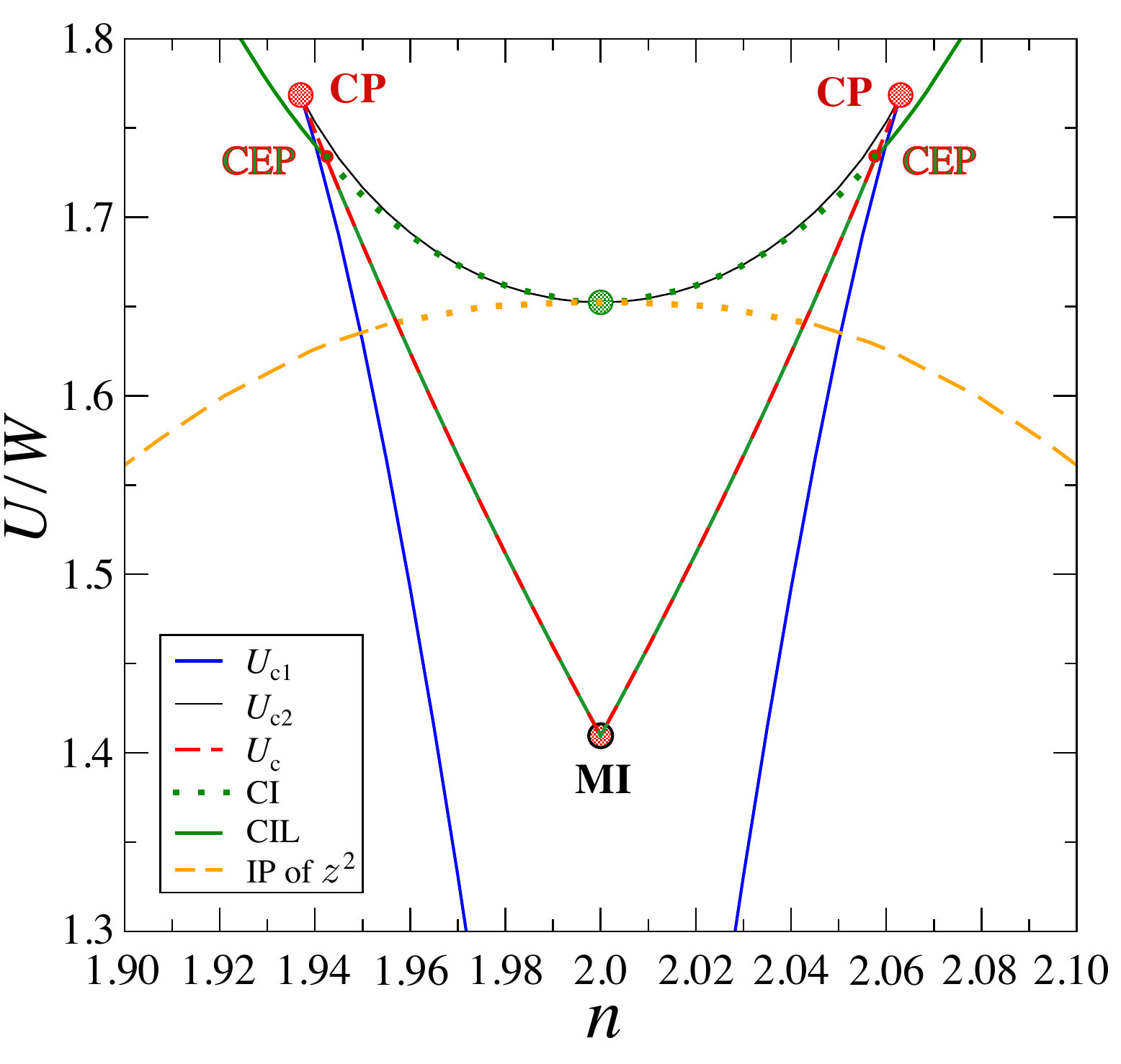}
\caption{Phase diagram for  
$J_{\rm H}=W/6$. The red dashed line separates a stable  metallic solution from a stable bad metal state at larger values of $U$. This line becomes red-green dashed where the charge instability coincides with that first order transition. $U_{c1}$ and $U_{c2}$ mark the boundary of the coexistence regime. 
The red circles locate the critical points ${\rm CP}=(n^*_{c}, U_{c}^*)$ and $(2-n^*_{c}, U_{c}^*)$. The green circle denotes $(n=2, U_{c2}(n=2))$. The charge instability line (CIL) merges with the $U_c$-line at the critical end point (CEP) marked  by the red-green dot (see the magnification of this regime in Fig.~\ref{Fig:pd_zoomed}). The first order
transition (red-green dashed line)  ends
at the MI-transition point, close to $(n=2,U/W=1.41)$. The green dots extend the charge instability (CI) into the metastable metallic state.
The orange dashed (dotted)  curve marks the inflection points of $z^2(n)$ in the metallic (metallic metastable) phase.}
\label{Fig:pd}
\end{figure}

\begin{figure}[!h]
\includegraphics[scale=0.60]{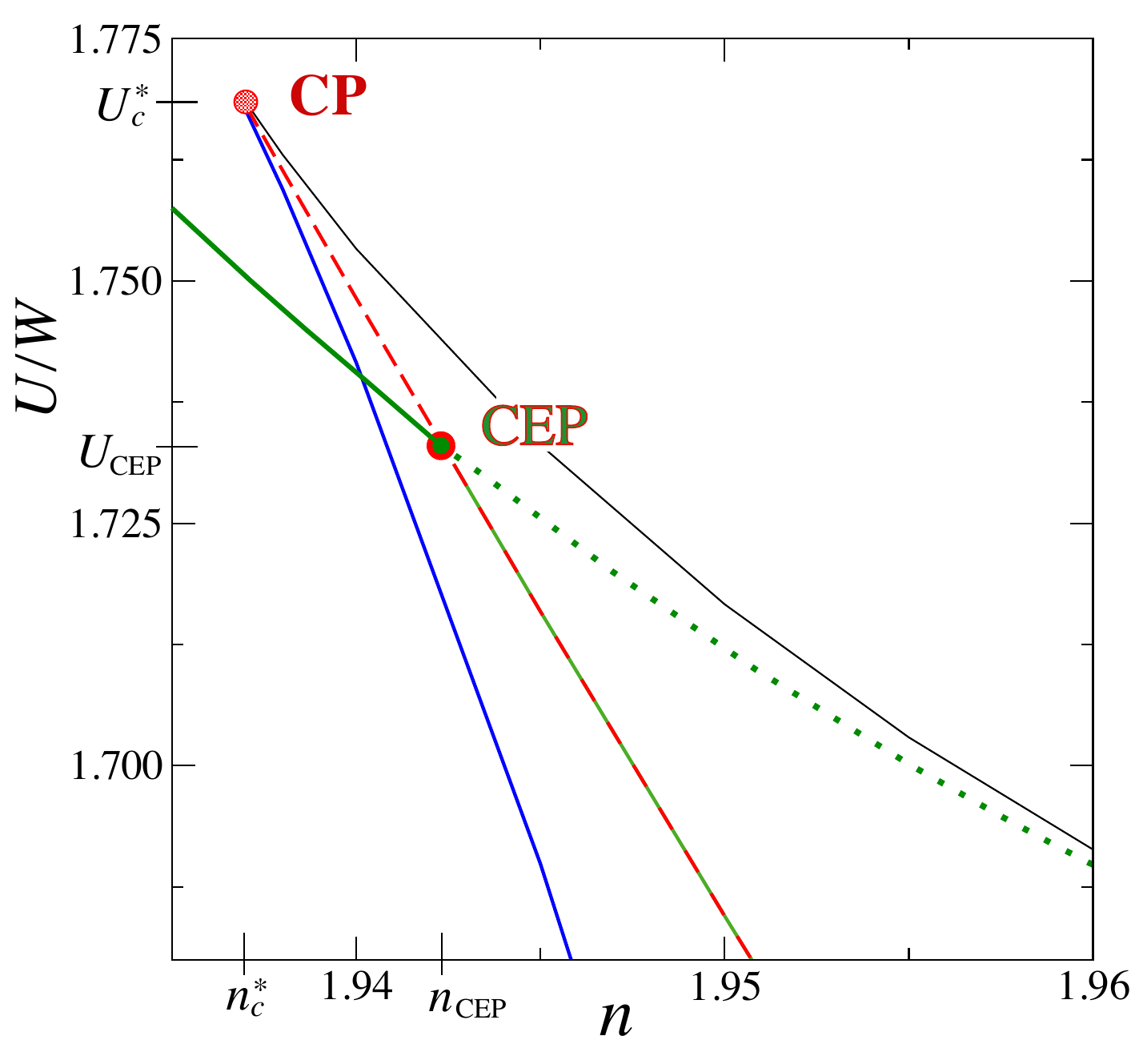}
\caption{Phase diagram of Fig.~\ref{Fig:pd} zoomed into filling range close to the critical end point (CEP).
}
\label{Fig:pd_zoomed}
\end{figure}

\begin{figure}[!h]
\includegraphics[scale=0.60]{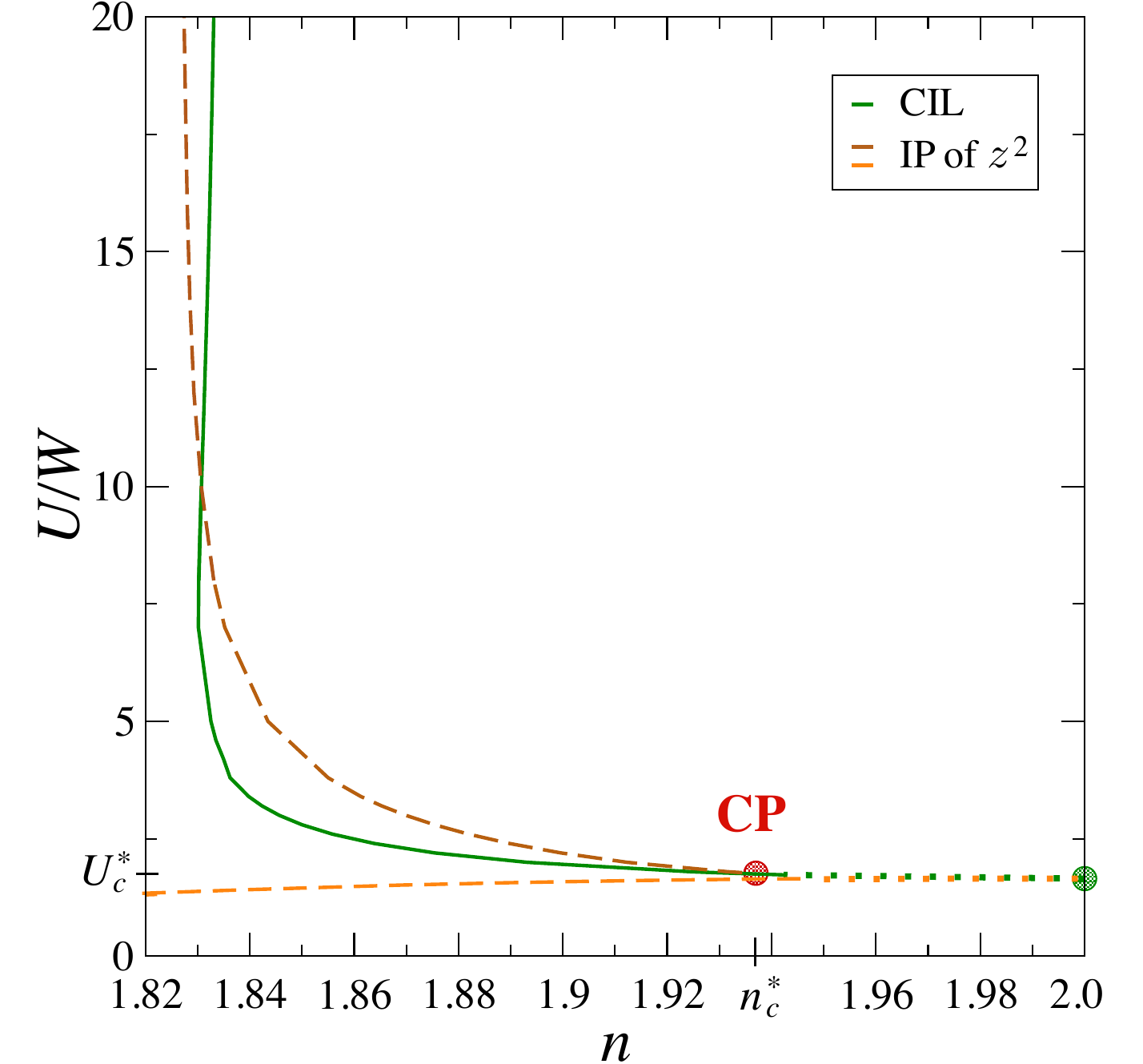}
\caption{Charge instability line (CIL) and lines  of inflection points of $z^2$ in the $(n,U)$
  phase diagram for $J_{\rm H}=W/6$. The red circle marks $(n^*_{c}, U_{c}^*)$, and the green
  circle $(n=2, U_{c2}(n=2))$. The dark-orange dashed line displays the inflection points 
  of $z^2(n)$ for $U$ above  $U_{c}^*$. The light-orange dashed line denotes the inflection points
  in the metallic regime (cf.\ to the diamonds in Fig.~\ref{Fig:z}) and the  light-orange dots 
  extend this line of inflection points into the metastable metallic state close to half filling.
  } 
\label{Fig:inst}
\end{figure}

The dependence of
the quasiparticle residue $z^2$ on charge carrier density and interaction
strength in the two-band model has been extensively investigated before
(see for example Refs.~\cite{Fresard1997, Fresard2001, Lechermann2007, Piefke2018}), not the least
because it is directly related to the inverse effective electronic mass. Of
particular interest is its behavior at the commensurate densities: for $n=1$
and $n=3$ it decreases smoothly with increasing interaction strength, and
vanishes at the metal-to-insulator transition. It is  a continuous transition, 
and bears much resemblance with the Brinkman-Rice transition \cite{Brinkman1970}. On the
contrary, it is first order for $n=2$ in its dependence on $U$~\cite{Fresard2001}.

Here we consider a fixed parameter value $J_{\rm H} = W/6$ and plot the quasiparticle residue
against filling for various values of $U$ (see Fig.~\ref{Fig:z}).  Its  behavior varies strongly with
the correlation strength: 
for  $U < U_{\rm MI} \simeq 1.41~W$ the quasiparticle residue smoothly depends on
filling below the Mott insulator transition (MI). As previously shown in Ref.~[\onlinecite{Fresard1997}], it displays a broad minimum at $n=2$ and decreases with increasing $U$.

The same seems to apply for $U$ up 
to $1.65~W$, but this is a fallacy. Indeed, a second solution starting from
$n=2$ with $z=0$ develops and is actually stabilized in a doping range around
half-filling that grows with increasing $U$. This marks a coexistence region
of the above described metallic state with an insulating-like doped Mott insulator
or ``bad metal'' state.
The bad metal state disappears below an $n$-dependent value
$U_{c1}(n)$ and the metallic state above a value $U_{c2}(n)$ (see blue and black
curves in Fig.~\ref{Fig:pd}, respectively, where the phase diagram is
presented, as well as in Figs.~\ref{Fig:pd_zoomed} and \ref{Fig:inst}.). The lowest value
of $U_{c2}$ is  $U_{c2}(n=2) \simeq 1.65~W$ for $J_{\rm H} = W/6$.
The metallic and the
insulating-like solutions are degenerate along the red dashed lines 
in Figs.~\ref{Fig:pd} and \ref{Fig:pd_zoomed}.
As a hallmark of this first order transition, the coexistence range of the two
solutions is rather limited in size and extends from $U =U_{\rm MI}$ at
half-filling to at most $n^*_{c} \simeq 1.937$
for $U^*_c \simeq 1.768~W$. 
Beyond it, both solutions turn indistinguishable and
accordingly smoothly connect. 

At the critical point CP, located at $(n^*_{c}$, $U^*_c)$, the residue $z^2$
possesses an inflection point in its density dependence, where its derivative
diverges. When further increasing $U$, there remains an inflection point,
where the magnitude of the slope steadily decreases (see the diamonds placed
on the continuous curves in Fig.~\ref{Fig:z} and the dark-orange dashed line
in Fig.~\ref{Fig:inst}). As the addressed jump of the quasiparticle residue
for $U<U_{c}^*$ transforms into an inflection point in its density dependence,
it is to be associated with a crossover. For $U < U_{\rm MI}$, besides the
stable metallic solution, there remains a solution arising from the Mott 
insulator. It is metastable and, therefore, it will not be addressed any
longer in the following.

In addition, we  display the charge instability line (CIL) of the metallic
solution as continuous green lines in Figs.~\ref{Fig:pd}, \ref{Fig:pd_zoomed}, and \ref{Fig:inst}. Along this line
the inverse electronic compressibility is zero. The CIL merges with the $U_c$-line
at $(n_{\rm CEP} \simeq 1.9423, U_{\rm CEP} \simeq 1.733~W)$ (see Fig.~\ref{Fig:pd_zoomed}). Below this value of $U$, the 
$U_c$-line not only represents the transition from metallic to bad-metal behavior but also
a discontinuity of $\kappa^{-1}$ (jump from positive to negative values of $\kappa^{-1}$).
The analysis of the charge instability will be presented in Sec.~\ref{sec:compressibility}.

\begin{figure*}[t]
\subfloat[]{
  \includegraphics[width=0.32\textwidth]{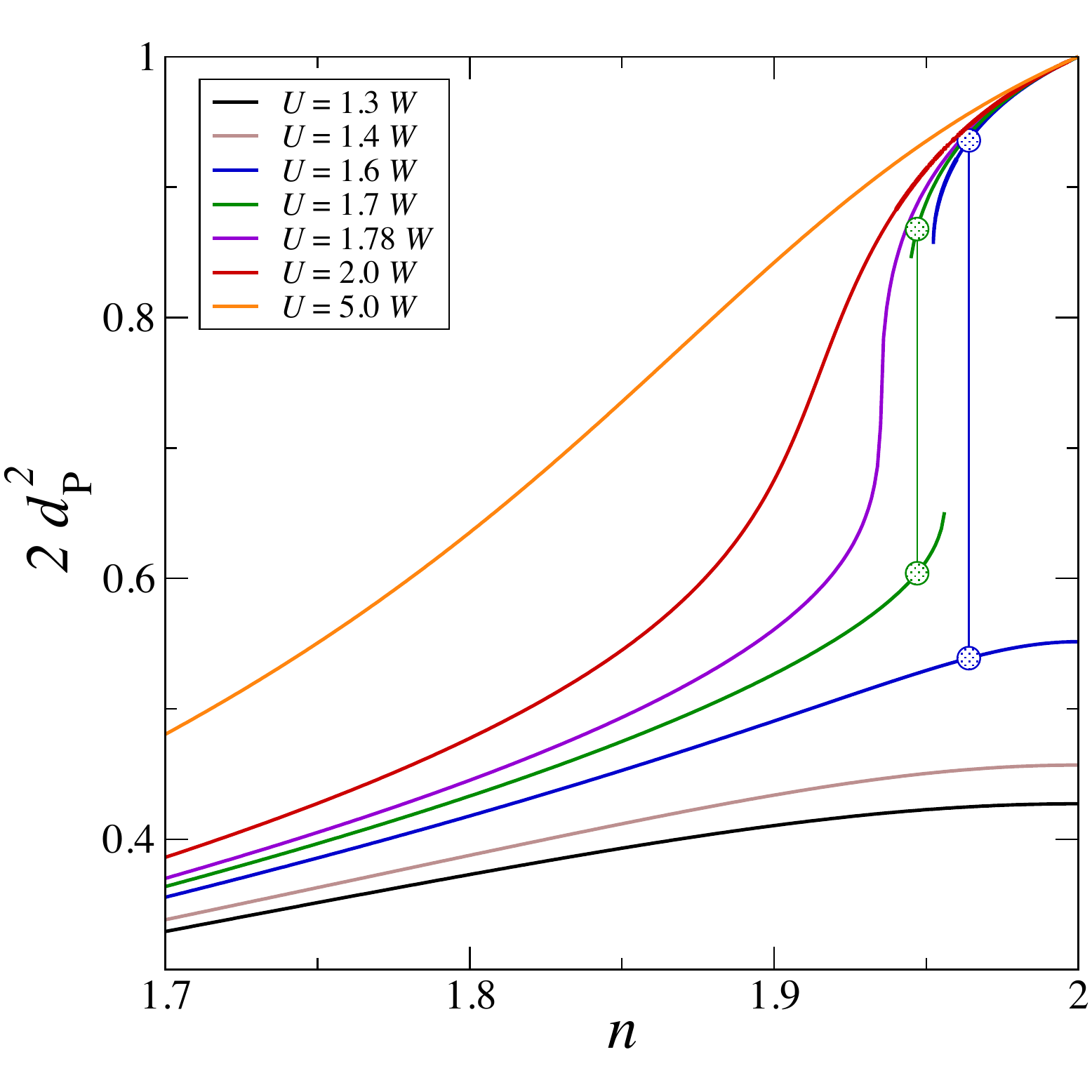}
}\hfil
\subfloat[]{
  \includegraphics[width=0.32\textwidth]{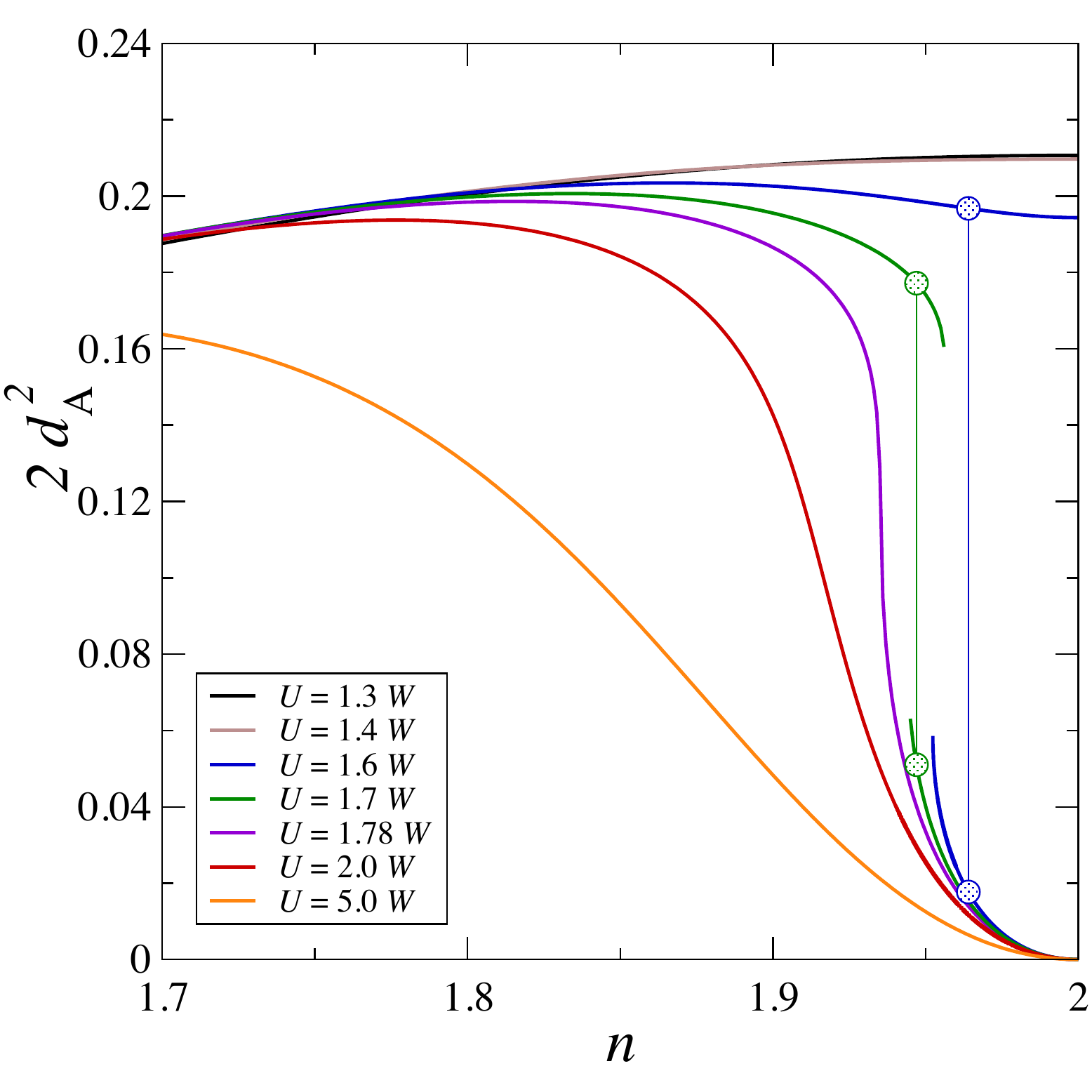}
}\hfil
\subfloat[]{
  \includegraphics[width=0.32\textwidth]{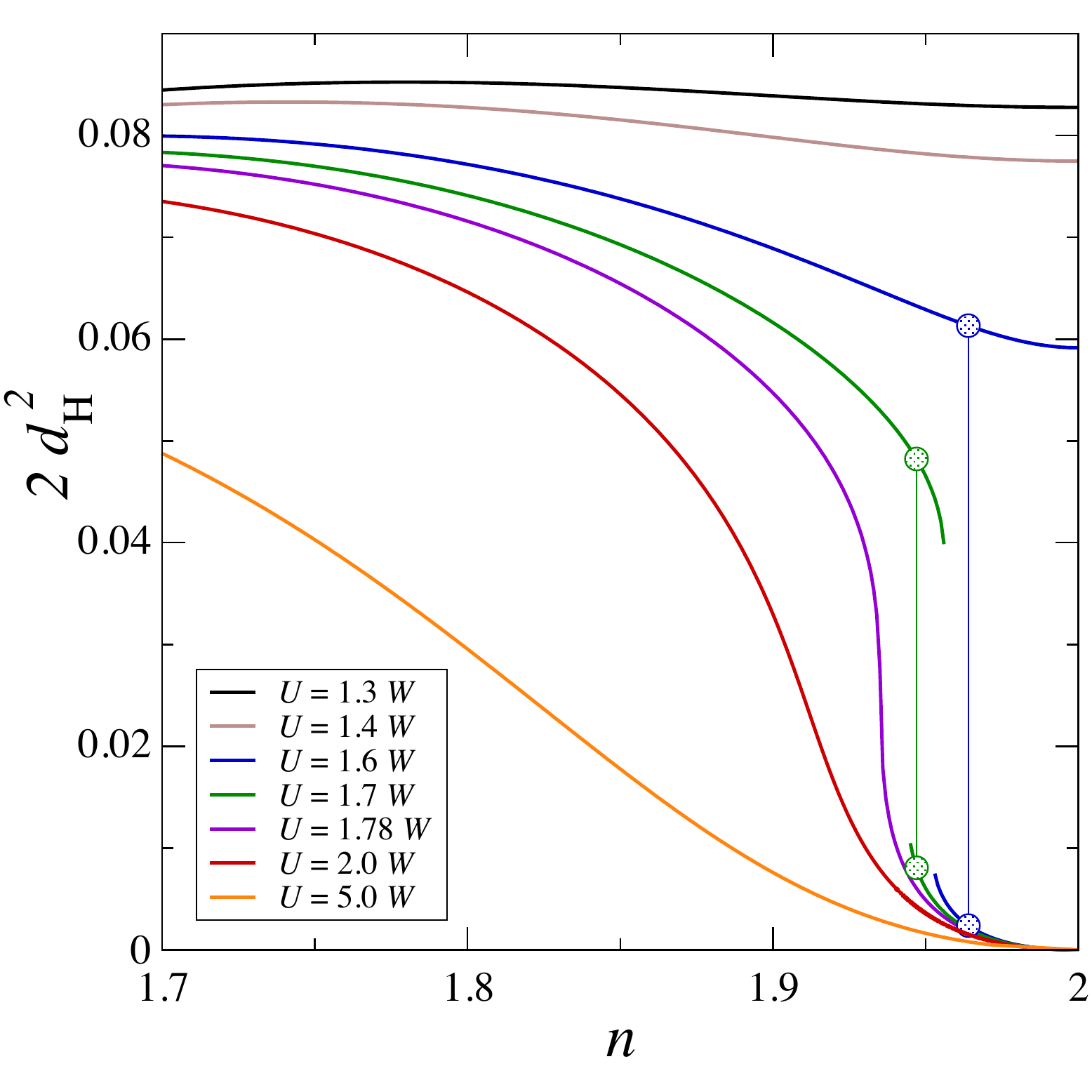}
}
\caption{
Filling-dependent expectation values of site double occupancies for $J_{\rm H} = W/6$: 
  (a) two parallel spins, 
  (b) two anti-parallel spins in different orbitals,
  (c) two anti-parallel spins in the same orbital.
The circles and the vertical thin lines characterize the first order
  transitions.  }
\label{Fig:ds}
\end{figure*}

\subsection{Slave boson fields}
\label{sec:SBF}

The slave boson expectation values represent  collective fields.
With the calculations performed at fixed $J_{\rm H}$ the collective fields
involving double occupancies markedly differ from one another in a broad
density range around half-filling, As shown in Fig.~\ref{Fig:ds}, the hierarchy 
$d_{\rm H}^2 < d_{\rm A}^2 < d_{\rm P}^2$ is always clearly obeyed, with the
exception of the Mott insulating phase where the first two vanish. The
critical point $(n^*_{c}, U_c^*)$ illuminates the density dependence of all
bosons; there, they all exhibit an inflection point with diverging derivative
with respect to $n$. For $ U > U_c^* $ inflection points remain, though 
the amplitude of the derivatives diminishes. On the other side, $ U < U_c^* $,
all boson expectation values jump at the first order transition whereas  a
smooth behavior is restored for $ U < U_c(n=2)$. 

\subsection{Compressibility}
\label{sec:compressibility}

The inverse electronic compressibility  is expressed through the derivative of the chemical 
potential $\mu$ with respect to the electronic density $\rho$ 
\begin{equation}
\kappa^{-1}= \rho^2 \,\frac{\partial\mu}{\partial\rho}
\end{equation}
where we consider the zero-temperature compressibility for constant volume.  The 
density in the two-dimensional electronic system
is trivially related to the filling through $n= a^2 \rho$ where $a$ is the lattice constant.
Alternatively, the inverse compressibility may be calculated directly from $F(n)$, which is the Legendre
transform of $\Omega(\mu)$, through $\kappa^{-1}=n^2 d^2 (F/N_{\rm L} a^2)/d^2 n$.

In this work we ascribe a continuous transition with a zero crossing in the inverse electronic compressibility 
$\kappa^{-1} (n)$ to a charge instability (see the green lines in Figs.~\ref{Fig:pd} and \ref{Fig:inst}). There the charge susceptibility $\kappa$ diverges.
On the other hand, a first order transition emerges if $\kappa^{-1} (n)$ changes discontinuously and the charge susceptibility stays finite. This discontinuity is tied to the metal to bad metal transition (see the red-green lines in Fig.~\ref{Fig:pd}). Only at singular points $(n_{\rm CEP},U_{\rm CEP})$, the inverse compressibility approaches zero from the low-filling side and jumps to a negative value (see the green-red point in Fig.~\ref{Fig:pd_zoomed}). There the charge instability line ends at the first order transition line. In analogy with
similar end points in thermodynamic phase transitions we denote this point as a ``critical end point'' (CEP).

We now analyze the formation of a negative compressibility state 
in few of the involved fermionic and bosonic  degrees of freedom.
The grand potential $\Omega$, Eq.~(\ref{eq:Omega}), as well as $F$ is made
of a fermionic contribution arising from the quasi-particles, and a bosonic
one, to which no coherence may be related. Accordingly, the full inverse
compressibility $\kappa^{-1}$ consists of a fermionic contribution 
$\kappa^{-1}_{\rm f}$, arising from the last two lines of Eq.~(\ref{eq:Omega}), and
a bosonic one $\kappa^{-1}_{\rm b}$, deduced from the first three lines of
Eq.~(\ref{eq:Omega}). Note that the last line is the kinetic energy for a fermionic system
and the second but last line contains a contribution to the constraint which relates the 
fermionic to the bosonic degrees of freedom.

\begin{figure*}[t]
\subfloat[]{
  \includegraphics[width=0.485\textwidth]{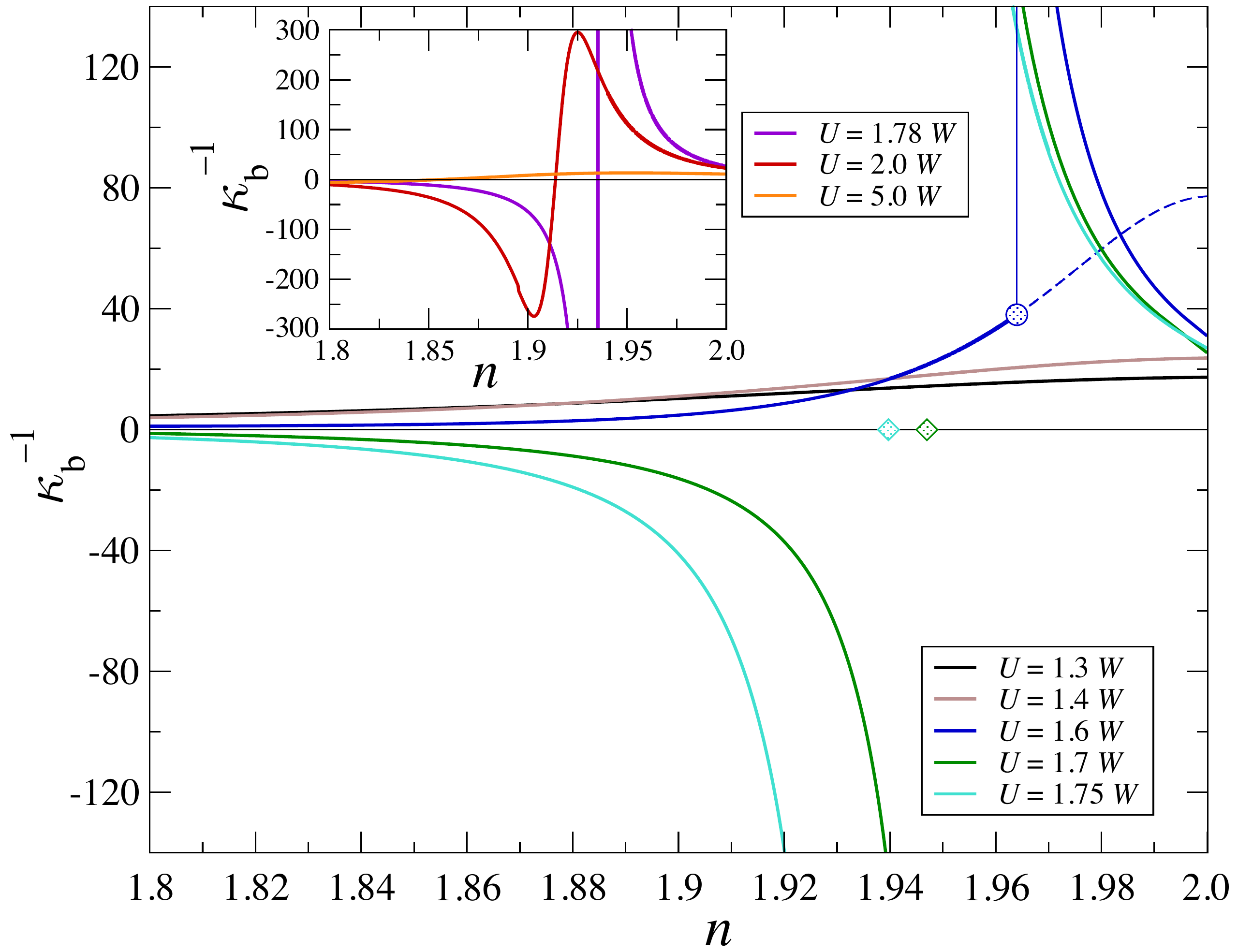}
}\hfil
\subfloat[]{
  \includegraphics[width=0.485\textwidth]{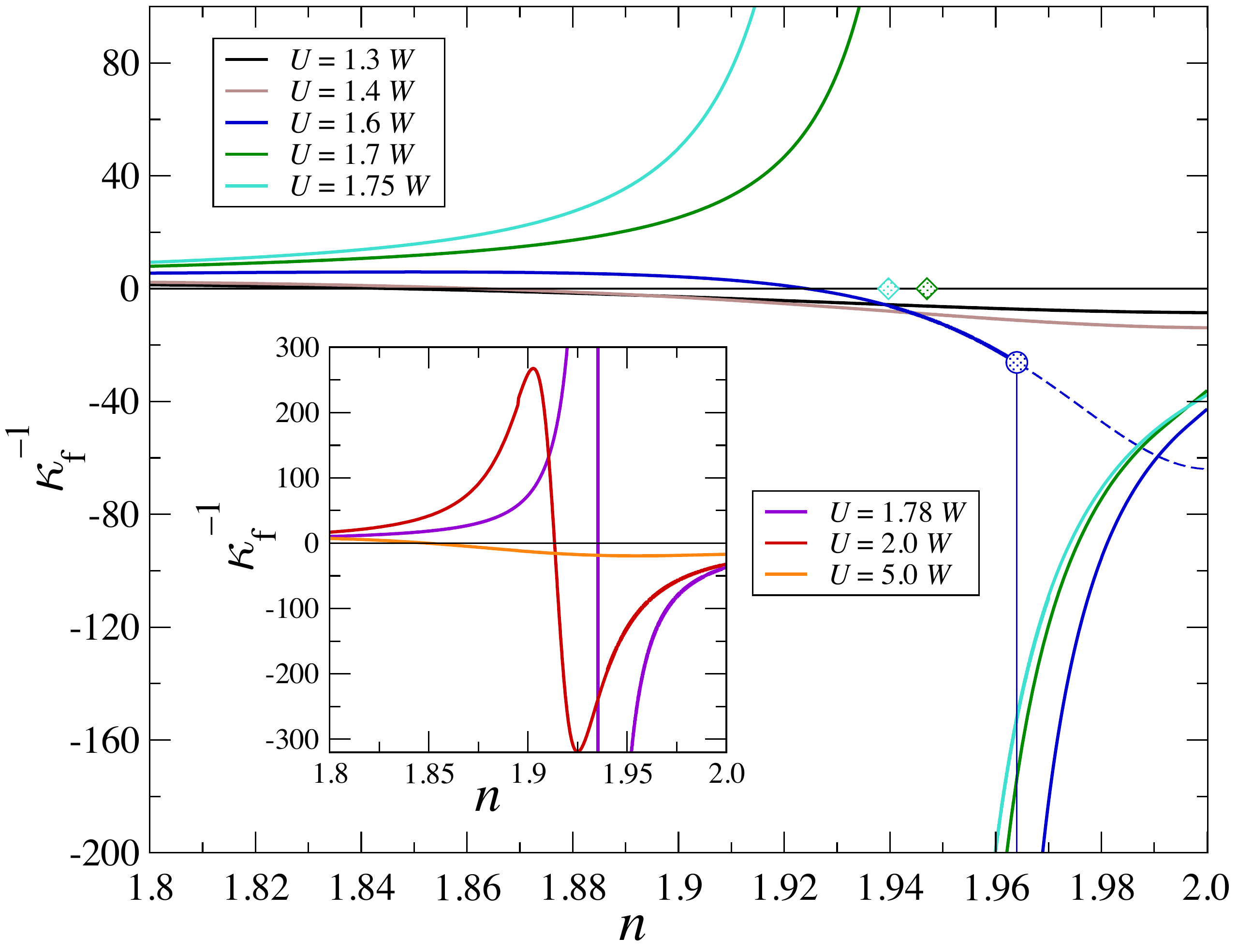}
}\caption{
 Bosonic (a) and fermionic (b) contribution to the inverse compressibility
  in dependence on filling $n$ for $J_{\rm H} = W/6$. 
  Here $\kappa_{\rm b,f}^{-1}$ is in units of $W/a^2$.
  The vertical blue line marks a first order transition for $U = 1.6\,  W$. The dashed blue line represents the metastable metallic case. The
  positions of the 
  first order transitions for $U = 1.7\, W$  and $1.75\,  W$ are schematized by diamonds. The transitions in the inset are
  continuous albeit the inverse compressibility for $U = 1.78\, W$ is not displayed completely. }
 \label{Fig:kappa_broad_Jp2}
\end{figure*}

The interaction distinctly influences $\kappa^{-1}_{\rm b}$ as may be deduced from Fig.~\ref{Fig:kappa_broad_Jp2}(a). 
For weak to moderate coupling
$U\lesssim 1.4~W$ all bosons display a comparatively weak density dependence and
this holds true for $\kappa^{-1}_{\rm b}$ as well. For $U$  above $U_{\rm MI}$, the bosonic contribution to $\kappa^{-1}$
is still positive but jumps to  a larger value close to half-filling in the bad metal state. Then, for $U>U_{c2}(2)$, the metastable
metallic state does not extend to $n=2$ and $\kappa^{-1}_{\rm b}$ is negative in a wide filling regime well below
half-filling before it jumps to a positive
value close to half-filling in the regime where the bad metal state is stabilized. Eventually, for $U\geq U_{c}^*$, $\kappa^{-1}_{\rm b}$ is continuous with  a minimum and a maximum
below and above the transition, respectively  (see inset of 
Fig.~\ref{Fig:kappa_broad_Jp2}(a)). 

 In order to relate these
findings to the bosonic fields one may rewrite the bosonic contribution $\Omega_{\rm b}$ to the grand
potential Eq.~(\ref{eq:Omega}) as
\begin{equation} \label{eq:omega_simple}  
\Omega_{\rm b}/N_{\rm L} = -d_{\rm P}^2 \left(U+ 4 J_{\rm H} \right) - 
           d_{\rm A}^2 \left(U +2J_{\rm H} \right) - 
           d_{\rm H}^2 \left(U - 2 J_{\rm H} \right)   
\end{equation}
This expression results from first neglecting the very small contributions
from the bosons $e$ and $\varpi$, and second to using the constraints to
express the boson $t$ in terms of the $d$-bosons. This additionally yields
terms proportional to $n-1$, but the latter do not contribute to the inverse
compressibility---as $\kappa^{-1}$ is the second derivative with respect to 
$n$---and they are not included in $\Omega_{\rm b}$
of Eq.~(\ref{eq:omega_simple})  for the following discussion of $\kappa^{-1}_{\rm b}$

Recalling the above hierarchy
among the $d$-bosons it turns out that the leading contribution to $\Omega_{\rm b}$
follows from the $d_{\rm P}$-boson. At $U=U_c^{*}$ the $d_{\rm P}$-boson
possesses an inflection point in its density dependence that is located at $n=n^*_{c}$. It
separates a density range where the density dependence of $d_{\rm P}^2$
is characterized by a positive curvature from a 
regime with a negative curvature (compare the purple curve in Fig.~\ref{Fig:ds}(a)). This sign
change of the curvature persists for larger $U$ values, which is reflected in
$\kappa^{-1}_{\rm b}$ (see inset of Fig.~\ref{Fig:kappa_broad_Jp2}). This is also true for $U<U_c^{*}$ 
but close to it---with the additional feature of a jump at the first order phase transition.

 For weak to moderate coupling $U< U_{\rm MI}$ the
curvature of the $d_{\rm P}$-boson contribution is negative in the entire presented density
range. As this is in fact the leading contribution to $\kappa^{-1}_{\rm b}$, it results necessarily 
in a positive bosonic compressibility. 

For  $U\gtrsim U_{\rm MI}$, the density dependence of
$d_{\rm P}^2$ displays a negative curvature in the considered regime close to half filling. In this case, $\kappa^{-1}_{\rm b}$
remains positive. However, well below half-filling the curvature switches its sign as seen for the blue
curve in Fig.~\ref{Fig:ds}(a)) even though the compressibility $\kappa_{\rm b}$ 
stays positive. 
This seeming inconsistency is resolved
by the observation that the contribution of the other two $d$-bosons in Eq.~(\ref{eq:omega_simple}) overcompensates the one
of $d_{\rm P}$ for this doping regime far from half filling.

The filling dependence of the fermionic contribution $\kappa^{-1}_{\rm f}$  to the inverse compressibility
is to a large extend opposite to that of the bosonic 
$\kappa^{-1}_{\rm b}$ (see Fig.~\ref{Fig:kappa_broad_Jp2}(b)).
Again, the qualitative behavior of such a contribution to the inverse compressibility can be derived from a single
dominant term, namely the second derivative of the quasiparticle residue with respect to filling.

In order to understand this we analyze the free energy
arising from the grand potential at $T=0$. The last two lines of
Eq.~(\ref{eq:Omega})---together with the Legendre transformation---lead to
a fermionic contribution $F_{\rm kin}$ to the free energy composed of the kinetic energy,
only. Since 
$F_{\rm kin}$ may be obtained analytically in the limit $t' \rightarrow
0$ with little impact on the numerical results, we adopt this
approximation below. In that case we obtain the kinetic energy per site as
\begin{equation}
F_{\rm kin}/N_{\rm L} = -2 z^2 \frac{W}{\pi} \cos{\Bigl(\frac{\pi \delta}{4}\Bigr)}
\label{eq:Fkin}
\end{equation}
where  doping $\delta \equiv n-2$ was introduced for convenience.
From Eq.~(\ref{eq:Fkin}) one may infer the leading contribution to $\kappa^{-1}_{\rm f}$ to be
given by 
\begin{equation}
a^2 \kappa^{-1}_{\rm f} \simeq -2 n^2 \frac{\partial^2 z^2}{\partial n^2} \frac{W}{\pi}
\cos{\Bigl(\frac{\pi \delta}{4}\Bigr)}\,.
\label{eq:kappafkin}
\end{equation}
Numerical tests prove Eq.~(\ref{eq:kappafkin}) to be a good
approximation. 

In the regime of weak to moderate coupling ($U \lesssim U_{\rm MI}$) the
effective mass $\sim 1/z^2$ weakly depends on filling, though featuring an
inflection point  (cf.\ the position of the diamonds in Fig.~\ref{Fig:z} and
the light-orange dashed line in Fig.~\ref{Fig:inst}) at which the curvature
switches from negative to positive when increasing the filling. Accordingly,
$\kappa_{\rm f}^{-1}$ takes comparatively small values, exhibits a sign
change, and its magnitude somewhat increases in the vicinity of half
filling. When intermediate coupling is considered in the metallic phase ($U
\simeq 1.6~W$) the same trends are followed, yet with a larger magnitude and,
close to half filling, with a jump of the fermionic inverse compressibility to
a larger negative value in the stable bad metal state.  For larger interaction
strength, ($U_{\rm c2}(2) \leq U \leq U_{\rm c}^*$) the inflection point of
$z^2$ vanishes. Instead, increasingly negative curvature is realized in the
entire metallic phase, while positive curvature characterizes the bad metal
phase. Note that the corresponding values taken by $\kappa_{\rm f}^{-1}$ close
to the discontinuity are too large to be displayed in
Fig.~\ref{Fig:kappa_broad_Jp2}(b). Once $U$ exceeds $U_{\rm c}^*$, the
inflection point of $z^2$ is restored, and so is the zero of 
$\kappa_{\rm f}^{-1}$. Let us stress that it remains a continuous function of
density that takes very large positive and negative values (see the purple
curve in the inset of Fig.~\ref{Fig:kappa_broad_Jp2}(b)).

Since $\kappa_{\rm b}^{-1}$ is mainly controlled by the $d_{\rm p}$-boson while
$\kappa_{\rm f}^{-1}$ is primarily ruled by the inverse effective mass, that
itself depends on the $d_{\rm P}$-boson, one may wonder why these two
contributions to $\kappa^{-1}$ do compete. To that aim we seek for an approximate
but reasonably accurate analytical form of ${\partial^2 z^2}/{\partial n^2}$ that
enters Eq.~(\ref{eq:kappafkin}). From the plethora of contributions to it (cf.~Eq.~(\ref{Eq:defz})
and the definitions in Eq.~(\ref{Eq:defsLR})), it turns
out that 
\begin{equation}\label{Eq:approxder_z2}
\frac{\partial^2 z^2}{\partial \,n^2} \simeq \frac{4 \left(d_{\rm P} +
    d_{\rm A}+ d_{\rm H}\right)^2 }{1-\left( \frac{n-2}{2} \right)^2}
\frac{\partial^2 (p+t)^2}{\partial n^2}
\end{equation}
is a good approximation. Here a numerical test shows that the term with the
second derivative of $d_{\rm P}^2$ is small as compared with  the retained 
term ~(\ref{Eq:approxder_z2}) (cf.~Figs.~\ref{Fig:p_Jp2}(a) and (b) to
Fig.~\ref{Fig:ds}(a)). 
Hence, while the sign of $\kappa^{-1}_{\rm b}$ is
essentially given by the curvature of $d_{\rm P}^2$, the one of
$\kappa^{-1}_{\rm f}$  follows from the curvature of $(p+t)^2$. 
Fig.~\ref{Fig:ds}(a) and Fig.~\ref{Fig:p_Jp2}(a) and (b) show that
they are opposite in sign in the largest part of the parameter space of
interest where they therefore compete.

\begin{figure}[t]
\includegraphics[scale=0.335]{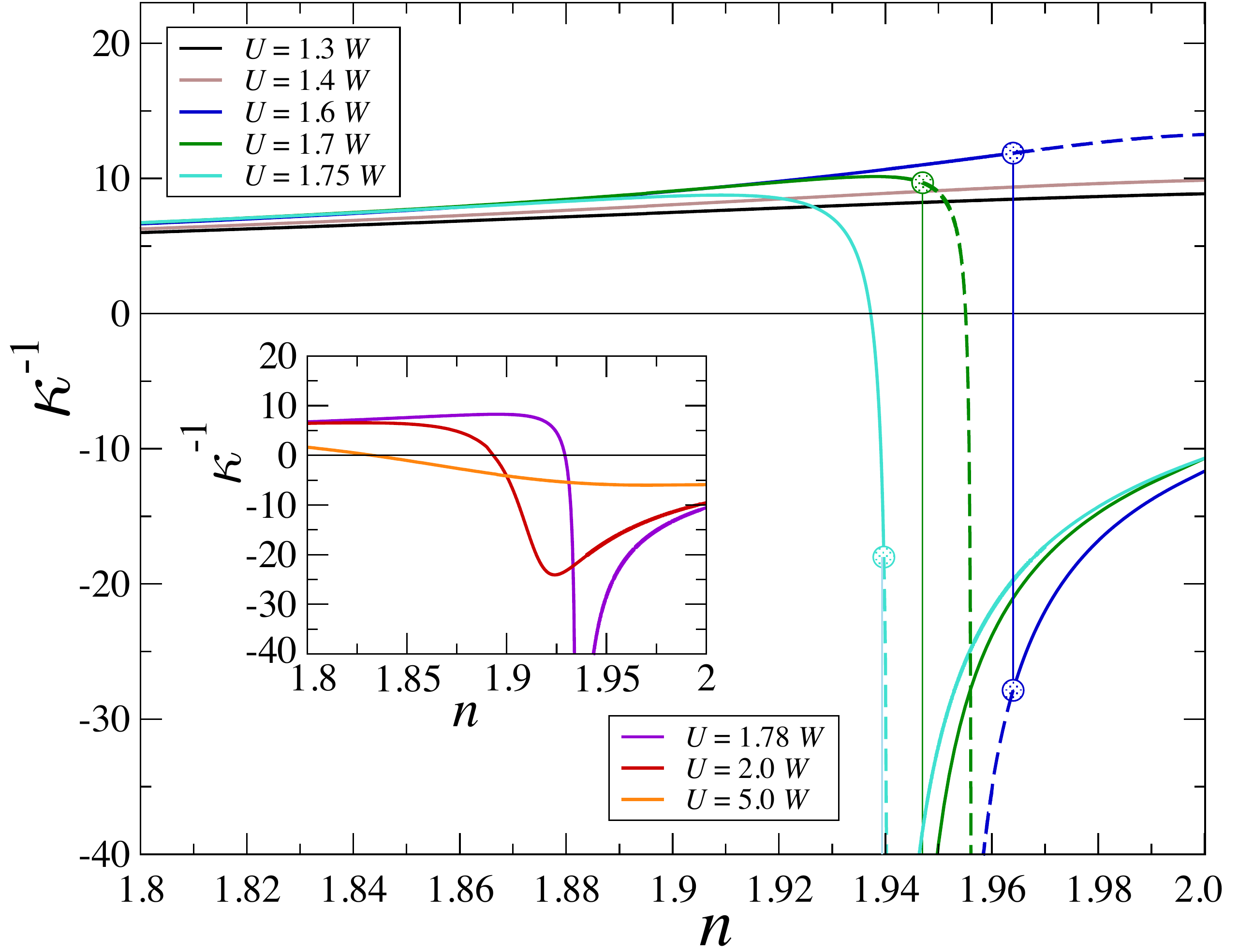}
\caption{Inverse compressibility in dependence on filling $n$ for $J_{\rm H} = W/6$. 
 Here $\kappa^{-1}$ is in units of $W/a^2$.
 The vertical lines mark a first order transition from the metallic to the bad metal state. 
 The dashed lines refer to the metastable states. }
\label{Fig:kappa_tot_broad_Jp2}
\end{figure}

The total inverse compressibility is shown in Fig.~\ref{Fig:kappa_tot_broad_Jp2}. 
The near cancellation of $\kappa^{-1}_{\rm b}$ and $\kappa^{-1}_{\rm f}$ is
particularly clear for the smallest densities, i.e., the largest doping. There, not only the magnitude
of $\kappa^{-1}$ is smaller than the larger of its components, but its
$U$-dependence is strongly suppressed. 
For weak to moderate $U$
($U \lesssim U_{\rm MI}$), and under an increase in density, the bosonic
contribution takes over in the entire presented density range, where
$\kappa^{-1}$ remains positive. 

For intermediate coupling, $U_{\rm MI} <U < U_{c}^*$, the sign of $\kappa^{-1}$ follows 
mostly the one of $\kappa^{-1}_{\rm f}$. However note that the strong increase of
$\kappa^{-1}_{\rm f}$ on the low-filling side of the discontinuity is nearly canceled by
$\kappa^{-1}_{\rm b}$. Therefore the charge instability (with negative compressibility) is
formed already in the stable metallic state (see the turquoise curve in Fig.~\ref{Fig:kappa_tot_broad_Jp2}).
That regime is identified in the phase diagram of Fig.~\ref{Fig:pd} where for fixed $U$ close to but below 
$U_{c}^*$ one first crosses the charge instability line (CIL) with increasing $n$ and only then observes for
slightly larger filling a transition to a bad metal state. This regime ends at
$(n_{\rm CEP} \simeq 1.9423, U_{\rm CEP} \simeq 1.733~W)$ where the CIL merges with 
the $U_c$-line (see the red-green point in the inset of Fig.~\ref{Fig:pd}).

For $U$-values above $U_{c}^*$ the inverse compressibility is continuous---as
are its partial contributions $\kappa^{-1}_{\rm f}$ and 
$\kappa^{-1}_{\rm b}$---and for $U/W$
below approximately 10 the CIL stays at the lower filling side with
respect to the line of inflection points (see Fig.~\ref{Fig:inst}). 
Again, it is the bosonic contribution which drives the
compressibility to negative values already before the fermionic contribution beyond
the inflection point enforces the negative compressibility state at smaller doping. 

\subsection{Capacitance of a heterostructure}
\label{sec:cap}

\begin{figure}[t]
\includegraphics[scale=0.58]{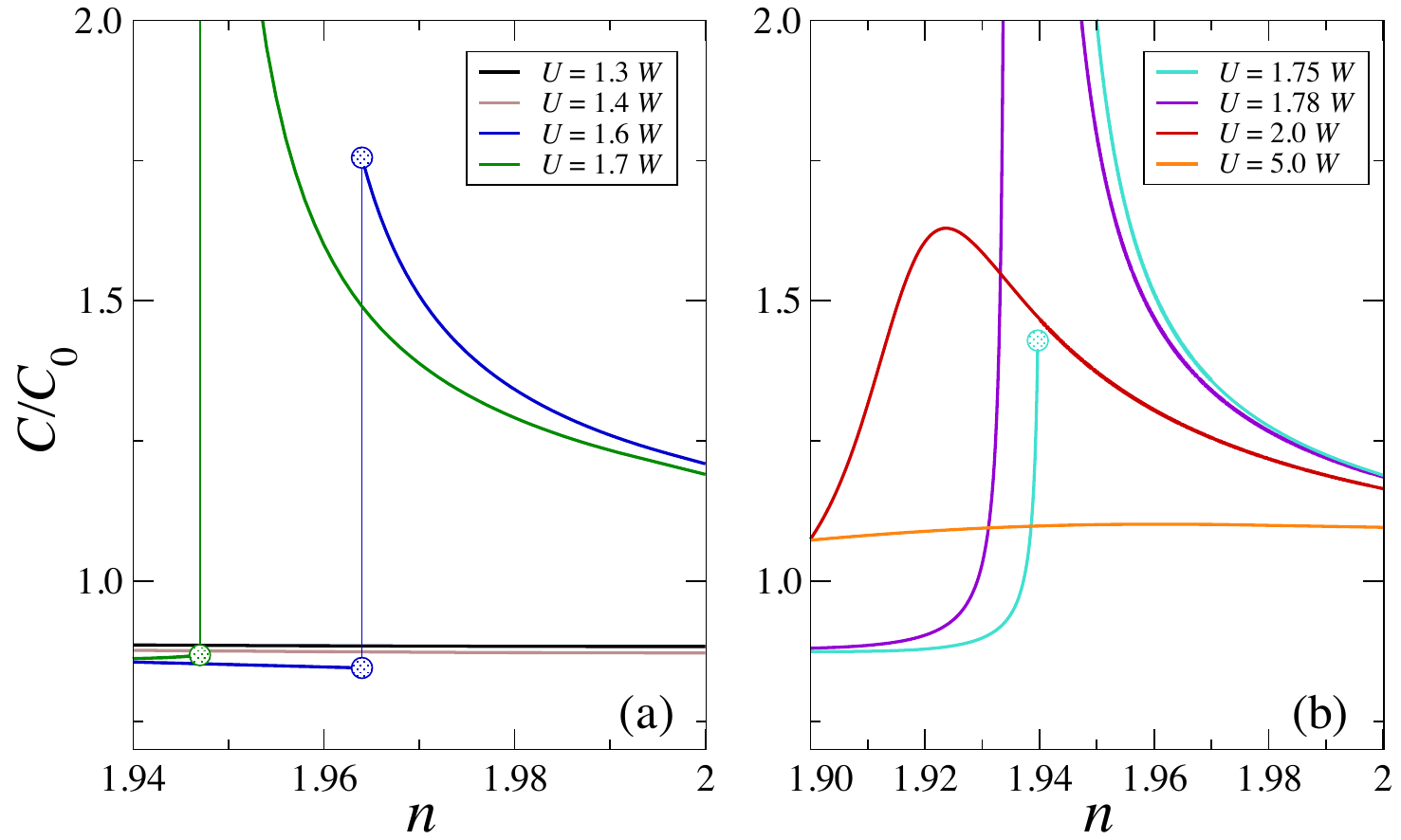}
\caption{Capacitance enhancement in dependence on filling $n$ for $J_{\rm H} = W/6$. (a) The weak coupling regime displays a continuous
filling dependence of $C/C_0$ with values below 1.0 whereas the
intermediate coupling range shows jumps of $C/C_0$ from less than 1 to values well beyond 1. The jumps of the capacitance are associated to
  the first  order metal to bad metal transition and are indicated by vertical lines. 
  (b) For  stronger coupling in the range from $U_{\rm CEP}$ to $U$ above $U_c^*$, the
  $C/C_0$-lines have two branches with a negative capacitance in between. For
  large coupling beyond $U_c^*$, the filling dependence of $C/C_0$ is again
  continuous. }
\label{Fig:Capacitance}
\end{figure}

Previous studies point out a tendency for the capacitance of heterostructures 
comprising strongly correlated electron systems to be larger than those with  
weakly interacting electron systems \cite{Steffen17,Berthod21}. Here, we consider a capacitor made of a polarizable
dielectric between two  electrodes as modeled by the current
two-band Hubbard model. In this simple set-up, the quantum corrections to the
inverse capacitance (see, e.g.\ Ref.~\cite{Kopp09}) are given by 
\begin{equation}\label{Eq:cap}
\frac{C_0}{C} = 1 + 2\frac{\varepsilon_0 \varepsilon a^2}{e^2 d } 
\frac{\partial \mu}{\partial n} \,.
\end{equation}
Here, $C_0=\varepsilon_0 \varepsilon A/d$ is the geometric capacitance of a capacitor
with two plates, $\varepsilon$ is the dielectric constant of the dielectric material
between the two electrodes, each of area $A$, and $d$ is the thickness of the dielectric. To be specific, we use the parameter
values $ {d}/{\varepsilon} = 4 a_{\rm B}$ (with $a_{\rm B}$ the Bohr radius), and the
lattice spacing $a$ is set to $6 a_{\rm B}$. The prefactor of ${\partial \mu}/{\partial n}$
in Eq.~(\ref{Eq:cap}) is then $2{\varepsilon_0 \varepsilon a^2}/(e^2 d) = 0.0526$~eV$^{-1}$.

As can be seen in Fig.~\ref{Fig:mu} the chemical potential steadily
grows with density in the largest part of the phase diagram.
This includes the weak coupling regime $U \leq U_{\rm MI}$ for all densities as well as
the moderate to strong coupling regime for large doping. In these regimes the kinetic
term rather acts to lower the capacitance. 

For moderate coupling in the range from  $U\simeq U_{\rm MI} $ to 
$ U \lesssim U_{\rm CEP}$ the metallic
state becomes unstable close to half filling and the compressibility jumps to negative
values in the bad metal state. Concomitantly, $C/C_0$ is pushed to a value well
above 1 which is easily understood from Eq.~(\ref{Eq:cap}) for the parameter regime 
where the right hand side (rhs) is still positive (cf.~Fig.~\ref{Fig:Capacitance}(a)).

For $ U\gtrsim U_{\rm CEP}$  the metallic state still persists in a small
doping range with negative compressibility and the rhs of Eq.~(\ref{Eq:cap}) is still positive
(see the turquoise curve in Fig.~\ref{Fig:kappa_tot_broad_Jp2} and the corresponding turquoise
curve for  $C/C_0$ in Fig.~\ref{Fig:Capacitance}(b)). The turquoise circle represents an end point
beyond which the capacitance is negative in a small doping range: When the bad metal state 
is stabilized at larger $n$, the inverse compressibility jumps to a more negative value. There 
the capacitance  $C$ becomes negative which signifies that the charging of the electrodes
changes (negative $C$ are not displayed in Fig.~\ref{Fig:Capacitance}). We do not investigate that charging instability further in this work (it was discussed in Ref.~\cite{Kopp09}). Eventually, with a slightly higher filling, the negative inverse compressibility
is again reduced sufficiently so that the rhs of Eq.~(\ref{Eq:cap}) becomes positive again and the second
branch of the (positive) capacitance curve close to half filling is observed. 

Eventually, for $U$ in the vicinity of $U_c^*$ (see the purple curves in Figs.~\ref{Fig:kappa_tot_broad_Jp2} and Fig.~\ref{Fig:Capacitance}(b)), the rhs of Eq.~(\ref{Eq:cap}) 
is zero twice in the regime of negative compressibility. Correspondingly, the capacitance diverges twice and it attains negative values around $n=1.94$. For even stronger coupling the dip in the inverse compressibility is less pronounced and the capacitance displays a broader maximum (see the red
and orange lines in Fig.~\ref{Fig:Capacitance}(b) for $U=2.0~W$ and $5.0~W$, respectively).

It is evident that, with the strong dependence of the capacitance on filling in the intermediate to strong coupling regime, switching capacitances through small electronic-density variations appears to be feasible.

\begin{figure}[!h]
\includegraphics[scale=0.48]{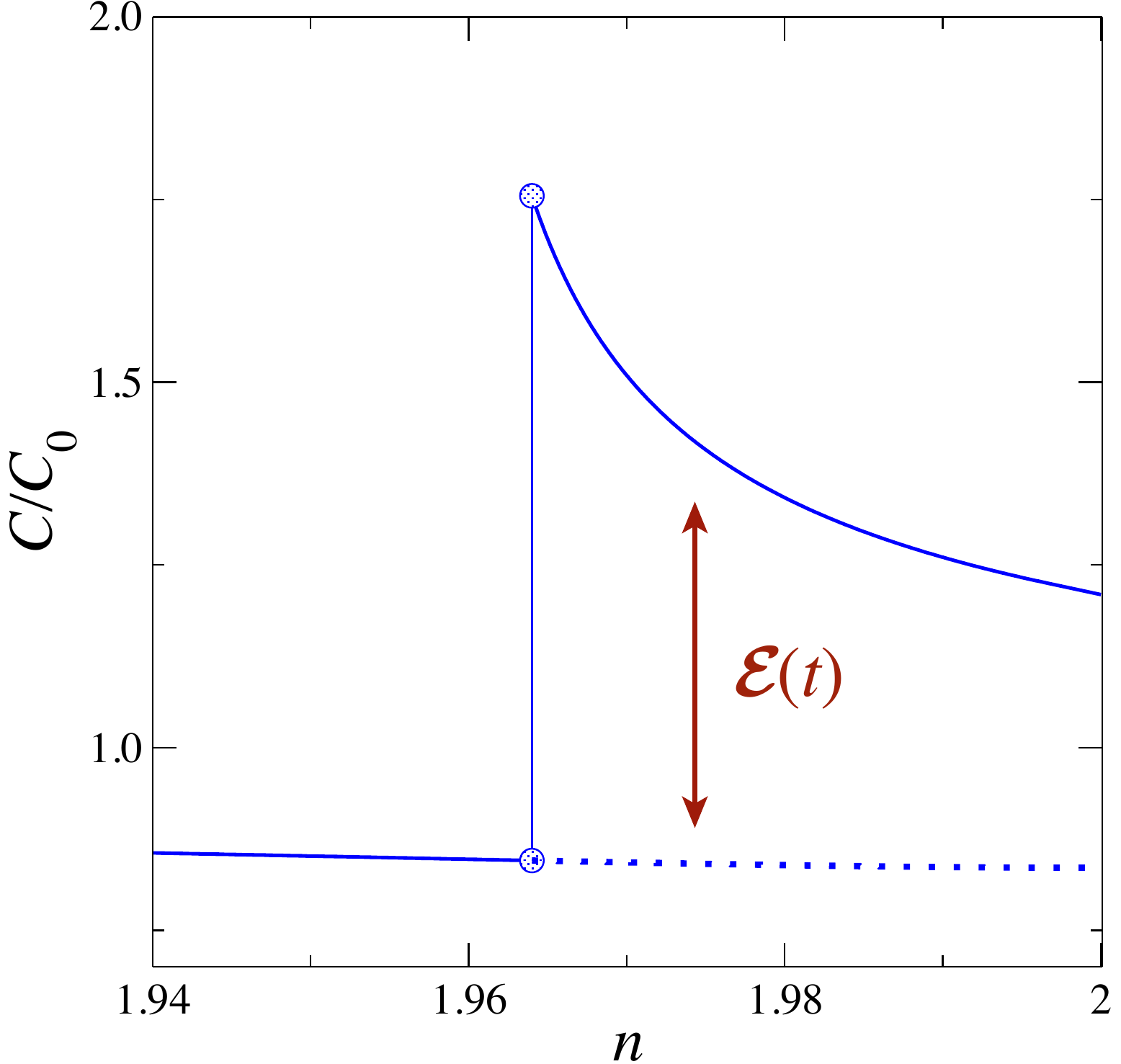}
\caption{Schematics of the capacitance switching induced by a short electric pulse ${\cal E}(t)$. The Hund's coupling is $J_{\rm H} = W/6$ and
$U=1.6~W$. Below approximately $n\simeq 1.965$, the continuous blue line represents the capacitance with electrodes in the stable metallic state, whereas above $n\simeq 1.965$ this line refers to the stable bad metal state.
The dotted blue line represents the metastable metallic case. 
The obtained effect is enhanced when approaching the transition, that is, when the energy difference between the two phases is smaller.     
}
\label{Fig:pulse_switching}
\end{figure}

Moreover, we suggest that with electric pulse switching between the high resistance Mott insulator and the low  resistance metallic state~\cite{Cario10} it is possible to switch between low and high capacitance in a corresponding device. This is indicated 
in Fig.~\ref{Fig:pulse_switching} for the capacitance transition with $U/W =1.6$.

\section{Blume-Emery-Griffiths approach}
\label{sec:BEG}

The phase diagram of the two-band Hubbard model close to half-filling is surprisingly intricate exposing a first-order and a continuous phase transition (Figs.~\ref{Fig:pd} and \ref{Fig:inst})---even though magnetic transitions are disregarded. The interpretation, however, is elusive as the slave-boson technique involves already seven bosonic fields in the paramagnetic state (at least four fields are relevant in the vicinity of half-filling) and their interplay jointly with the fermions is to be understood.

In the spirit of the Ising lattice-gas formulation of the liquid-gas transition we intend to mimic the bosonic fields by Ising-like pseudospins. The procedure builds on the assumption that it is not unreasonable to represent bosonic fields by classical fields and that the metal to bad metal transition is controlled by the bosonic degrees of freedom.
For this purpose we  simplify the formalism with exclusive focus on these transitions. We introduce a Blume-Emery-Griffiths (BEG) model~\cite{BEG1971} for the pseudospin degrees of freedom to better capture the machinery of the transition rather than gain quantitative results.

Such a simplification will not pave the way to reproduce the Mott transition or magnetic transitions close to half-filling. However it will generate qualitatively similar results as those in the previous section and thereby allows to understand the addressed transitions eventually in the more comprehensive framework of slave-boson theory. 

\subsection{BEG model and relation to the slave-boson representation}
\label{subsec:BEGmodel}

The basic idea is to interpret the bosonic degrees of freedom in terms of classical Ising-fields (pseudospins) which are controlled by various couplings. Foremost, there is  a Zeeman-like coupling term which provides the energetical splitting between different configurations of doubly occupied sites, so that the Hund's coupling $J_{\rm H}$ becomes a pseudo magnetic field for the Ising fields. Then there is for sufficiently strong interaction $U$ an effective nearest-neighbor exchange between orbital states  of doubly occupied sites that is translated into an Ising-type nearest-neighbor coupling of the pseudospins. 

The feedback of the Ising pseudo spins to the fermionic subsystem controls the kinetics of the fermions. It is the effective mass or rather the quasiparticle residue 
$z^2$ (see Eq.~(\ref{Eq:defz})) through which the bosonic fields affect the kinetic fermionic term. We will use the  dependency of $z^2$ on the bosonic fields, now for the dependence of $z^2$ on the classical Ising fields. 

Yet there is a second (reverse) feedback mechanism: the fermionic degrees of
freedom are expected to control the bosonic fields, that is, the Ising fields
in this approach. With the fermions coupled to the pseudospins, the latter
must necessarily fluctuate, even though they are introduced as classical
fields. Although this is a rather crude approximation we introduce the
effective bandwidth of the fermions as a soft energy cut-off for the
fluctuations of the pseudo spins by implementation of this energy cut-off as
an effective temperature for the pseudospins. In fact, we will see that this
approximation reproduces the slave-boson results qualitatively.

In detail we now proceed as follows:
In order to keep the number of pseudospin components minimal we only consider the fields related to the three doubly occupied states and the field representing the singly occupied sites. This will be sufficient for an intermediate coupling regime below half-filling (but above quarter filling). Later we will address the shortcomings of this reduction of degrees of freedom. Moreover one of the three fields representing doubly occupied sites may be related to the further fields through a constraint (see Eq.~(\ref{eq:relds})). Consequently we consider a spin-one Ising Hamiltonian for the pseudospins $S_i$. We identify $S_i=1$ with the spin-parallel occupation of the two orbitals on a site $i$, i.e.\ with $d^2_{\rm P}$, and $S_i=-1$ with the spin-antiparallel occupation of the two orbitals, i.e.\ with $d^2_{\rm A}$. The singly occupied sites are represented by $S_i=0$, which relates to the slave boson field $p^2$ on that site.

For arbitrary nearest-neighbor contributions the pseudospin representation of the bosonic degrees of freedom leads to a generalized form of the Blume-Emery-Griffiths (BEG) model~\cite{Saito1981}. We find that an antiferromagnetic nearest-neighbor (bilinear) Ising coupling  is consistent with the slave-boson results;  we will also provide a discussion for the choice of valid BEG-parameter regimes in Appendix~\ref{app_BEG_parameters}.

The Hamiltonian for the generalized BEG-model has the following structure for Ising spins on  $N_{\rm L}$  sites with nearest-neighbor coupling:
\begin{align}
{\cal H} = -{\cal J} \sum_{\langle i,j \rangle}  S_i S_j  &-{\cal K} \sum_{\langle i,j \rangle}S_i^2 S_j ^2 + \Delta \sum_{i=1}^{N_{\rm L}} S_i^2 -h \sum_{i=1}^{N_{\rm L}} S_i \nonumber \\
& - {\cal L}  \sum_{\langle i,j \rangle}( S_i S_j ^2 + S_i^2 S_j ) + E_0 \label{BEG-H}
\end{align}
which is the most general Hamiltonian for three classical states per site and nearest-neighbor coupling~\cite{Saito1981}. Here ${\cal J}$ is the coupling which controls ferromagnetic ${\cal J}>0$ or antiferromagnetic ${\cal J}<0$ correlations of the Ising pseudospins, that is, in the language of the two-band Hubbard model, it favors double occupancy  with the same or different orbital states on neighboring sites, respectively. The ``magnetic field`` $h$ aligns the pseudospins and corresponds to the Hund's  coupling: $h=J_{\rm H}/2$  (see Appendix~\ref{app_BEG_parameters}). The parameter $-\Delta$  in Eq.~(\ref{BEG-H}) controls the number of sites with zero pseudospin and is related to $\mu$, the chemical potential. Therefore we refer to it as the chemical potential related to the pseudospin particles. It will be fixed by the filling $n$.

The coupling ${\cal L} $ is to be included if the nearest-neighbor interaction
strength in $d_{\rm P}-d_{\rm P}$ configurations and the strength in $d_{\rm
  A}-d_{\rm A}$ configurations is not equal (see Appendix~\ref{app_BEG_parameters}). Here we
refer to a $d_{\rm P/A}-d_{\rm P/A}$ configuration when two neighboring sites
are both occupied by a $d_{\rm P/A}$ boson. Obviously, such terms with finite
${\cal L} $ denote in mean-field theory a shift of both $h$ and $\Delta$
proportional to ${\cal L} \langle S_i^2 \rangle$ and to $-{\cal L}  \langle
S_i \rangle$, respectively. In that respect, the coupling ${\cal L} $ is not
relevant for the existence of the discussed transition although it
renormalizes the other couplings.  In this section  we only consider the
BEG-model~\cite{BEG1971} where ${\cal L}=0$ and address finite ${\cal L} $ in
Appendix~\ref{app_BEG_parameters}.  

As to the fermionic dispersion, Eq.~(\ref{eq:Ek}), the $z^2$-factor is reduced to
\begin{equation}
  z^2 = \frac{4p^2\,(d_{\rm P}+d_{\rm A}+d_{\rm H})^2}
  {1-\left(\frac{2-n}{2} \right)^2}
\label{eq:z2}  
\end{equation}
in the approach with only four bosonic degrees of freedom (cf.\ Eq.~(\ref{Eq:approxder_z2}), where the triple occupation was included for a better
quantitative estimate of the compressibility). We stay below half-filling ($n<2$) because the corresponding results above half-filling may be derived directly from particle-hole symmetry, and we introduce $\delta = 2-n$ as the doping parameter. The relative number of singly occupied sites is $4 p^2 =\delta$ which in BEG is the relative number of zero-spin sites. Here, the factor 4 accounts for the two-spin directions and the two orbitals per site. The bosonic field $d_{\rm H}^2 $ is fixed by the relation~(\ref{eq:relds}). As in BEG the $d_{\rm P}$ and $d_{\rm A}$ configurations are assigned to spin $1$ and spin $-1$, respectively, one immediately identifies
\begin{subequations}
\begin{align}\langle S_i\rangle&\,= \,2 d_{\rm P}^2 - 2 d_{\rm A}^2\,\equiv m\,\label{eq:Si}\\
\langle S_i^2\rangle&\,=\, 2d_{\rm P}^2 + 2d_{\rm A}^2\,\equiv q\,\label{eq:Si2}
\end{align}
\end{subequations}
where we introduced the standard BEG-notation for the mean-field values of $S_i$ and $S_i^2$, that is, $m$ and $q$, viz.\ pseudospin magnetization and relative number of sites with pseudospin 1. Filling is expressed by $ n = 1+ 2d_{\rm P}^2 + 2d_{\rm A}^2  + 2d_{\rm H}^2$ if only four bosonic fields are considered and this expression may be rewritten as
\begin{equation}
q = 1-\delta -  2d_{\rm H}^2.\label{eq:q-dH}
\end{equation}
To include the field  $d_{\rm H}^2 $  through a constraint is consistent with the counting, however the sites with $d_{\rm H}$-configuration are not represented by a proper term in the Hamiltonian. This approach is justified if  the number of such sites, that is $d_{\rm H}^2 $, is much smaller than $d_{\rm P}^2 $, $d_{\rm A}^2 $ and doping $\delta$ which is true close to the considered transition (see the results below). One may introduce an on-site energy for the sites with $d_{\rm H}$-configuration but this accounts just for a shift of the chemical potential $\Delta$ and of the coupling constant ${\cal K}$  which does not affect our  mean-field results qualitatively.

It is straightforward to derive from Eqs.~(\ref{eq:relds}), (\ref{eq:z2}) and (\ref{eq:q-dH}) the following expression
\begin{widetext}
\begin{equation}
z^2(m,q)\,=\, \frac{1}{2} \frac{(1-q^2 -2d_{\rm H}^2)(q+2d_{\rm H}^2+ 
\sqrt{q^2-m^2}+2d_{\rm H}\, (\sqrt{q+m}+\sqrt{q-m}))}{1-\frac{1}{4}
(1-q-2d_{\rm H}^2)^2} 
\label{eq:z2qm}
\end{equation}
\end{widetext}
where 
\begin{equation}
d_{\rm H}^2(m,q)\,=\, \frac{1}{4}\, \frac{q^2-m^2}{13q+5m-12\sqrt{q^2-m^2}} \,.\label{eq:dH2}
\end{equation}
The variables $q$ and $m$ are taken from the mean-field solutions of the BEG model. The filling $n$ is found parametrically from 
\begin{equation}
n(m,q)\,=\,1+q +2d_{\rm H}^2(m,q)\label{eq:filling}
\end{equation}
which is equivalent to Eq.~(\ref{eq:q-dH}).

It is convenient to determine the upper and lower bounds for $z^2$:
\begin{equation}
 \frac{1}{2} \, f(\delta)
 \leq z^2 \leq \frac{3}{2}\, f(\delta) \quad {\rm with} \;\; f(\delta)=\frac{\delta(1-\delta)}{1-\frac{1}{4}\delta^2}
 \label{eq:z2-boundaries}
\end{equation}
which is valid for the considered case of four distinct on-site states. The lower bound is derived from full polarization, that is $m=q$  which implies $d_{\rm A}^2 =0=d_{\rm H}^2$ and $2 d_{\rm P}^2 = 1-\delta$. The upper bound is the ``non-magnetic'' state with $m=0$ which implies $2d_{\rm A^2} = 2 d_{\rm P}^2=2 d_{\rm H}^2 = (1/3)(1-\delta)$.

The BEG mean-field free energy $F$ of the paramagnetic state in the presence of finite field $h$ is (see Refs.~\cite{BEG1971,Saito1981}):
\begin{align}
F(T,h,\Delta)/N_{\rm L}=&\frac{1}{2} \zeta {\cal J} m^2 + \frac{1}{2}  \zeta {\cal K} q^2\nonumber\\
& +\! k_{\rm B} T \ln\bigl[1\!+ 2\, e^{-\frac{(\Delta-\zeta  {\cal K} q)}{ k_{\rm B} T}} 
\!\cosh\frac{\zeta  {\cal J} m+h}{k_{\rm B} T}\bigr] \label{eq:BEG-F}
\end{align}
where $ \zeta$ is the number of nearest-neighbor sites. We may cast the mean-field equations $\partial F/\partial m =0$ and $\partial F/\partial q =0$ into the form:
\begin{align}
 h&=-\zeta {\cal J} m + \frac{k_{\rm B} T}{2}\ln\frac{q+m}{q-m}
\label{eq:MF-h}\\
 \Delta&=\zeta {\cal K} q - \frac{k_{\rm B} T}{2}\ln\frac{q^2-m^2}{4(1-q)^2}
 \label{eq:MF-delta}
\end{align}

For such a classical Ising-type model a zero-temperature evaluation produces phase transitions where $m$ and $q$ change discontinuously (see, e.g., Fig.~2 in Ref.~\onlinecite{Saito1981}). This is not necessarily expected for the concomitant bosonic fields $d_{\rm P}$ and $d_{\rm A}$ (see Figs.~\ref{Fig:ds}(a) and (b) above) that are related to $m$ and $q$ through the identification (\ref{eq:MF-h}) and (\ref{eq:MF-delta}). These bosonic fields are in fact enslaved by the fermionic degrees of freedom and the challenge is then to allow for a control of the pseudospins through the fermions, at least approximately. We achieve this through a feedback mechanism where we assume that the temperature of the pseudospin BEG-system is an effective temperature which is proportional to some power $\alpha$ of the fermionic bandwidth: $k_{\rm B} T_{\rm eff} = g_{\rm fb}\, z^{2\alpha}$. Here $g_{\rm fb}$ is a (fermion-boson) coupling constant which however depends on $\alpha$ and will be discussed below. 

The  excitations of the bosonic system involve  fermionic Greens functions (or rather spectral functions) which are weighted by $z^2$. As virtual particle-hole excitations couple to the bosonic (pseudospin) degrees of freedom, one may assume in view of a perturbative approach that $\alpha=2$ is a suitable choice. For strong coupling, that is $t\ll U$, this may not be valid anymore and it may be argued that the excitations exist in an energy window given by the bandwidth $z^2 W$. Accordingly, one would then rather switch to $\alpha=1$ with increasing coupling.
So far there is no microscopic scheme how to determine $T_{\rm eff}$ and $\alpha$ \cite{comment1}. We find that the choice $\alpha=1$ does not produce  a discontinuous transition. Here we investigate the case with $\alpha=2$ which allows to reproduce the slave-boson results qualitatively when ${\cal J}$ is chosen appropriately. Consequently we introduce the effective temperature of the pseudospin system through
\begin{equation}
k_{\rm B} T_{\rm eff} = g_{\rm fb}\cdot z^{4} \label{eq:Teff}
\end{equation}
where $z^2$ is a function of $m$ and $q$ (see Eq.~(\ref{eq:z2qm})) and the fermion-boson coupling $g_{\rm fb}$ is chosen such that we recover the position of the jump or inflection point of $z^2$ in dependence on filling $n$ of the slave-boson results. This filling is denoted by $n_0$. We emphasize that this pseudospin approach to the bosonic fields is necessarily a phenomenological approach where the ``temperature profile'', that is, the dependence of the pseudospin temperature $T_{\rm eff}$ on $n$ (expressed through $q(n)$ and $m(n)$) is controlled by the strength of the coupling parameter $g_{\rm fb}$ and the effectiveness of the feedback mechanism, determined by the exponent  $\alpha$.

It is evident that in the limit of half-filling, $T_{\rm eff}$ converges to zero as $z^2$ approaches zero. This reproduces the correct limits of the fields $d_{\rm P}$, $d_{\rm A}$ and $d_{\rm H}$ but we do not consider this pseudospin approach as appropriate to discuss the Mott transition. We rather discuss the results below half-filling where our approach provides a transition to a state with negative compressibility in line with the slave-boson results.

\subsection{Results from the BEG approach and interpretation}
\label{subsec:BEGresults}
The procedure to calculate  $z^2$ in dependence on $n$ is as follows: We gain $m$ from the mean-field equation~(\ref{eq:MF-h}) for given $q$, whereby we replace the temperature by the effective temperature $T_{\rm eff}$ from Eq.~(\ref{eq:Teff}). Then we  use relations~(\ref{eq:z2qm}) and (\ref{eq:dH2}) to determine $z^2$. We plot $z^2$ in dependence on filling $n$ which is given by Eq.~(\ref{eq:filling}). 

\begin{figure}[!t]
\includegraphics[scale=0.62]{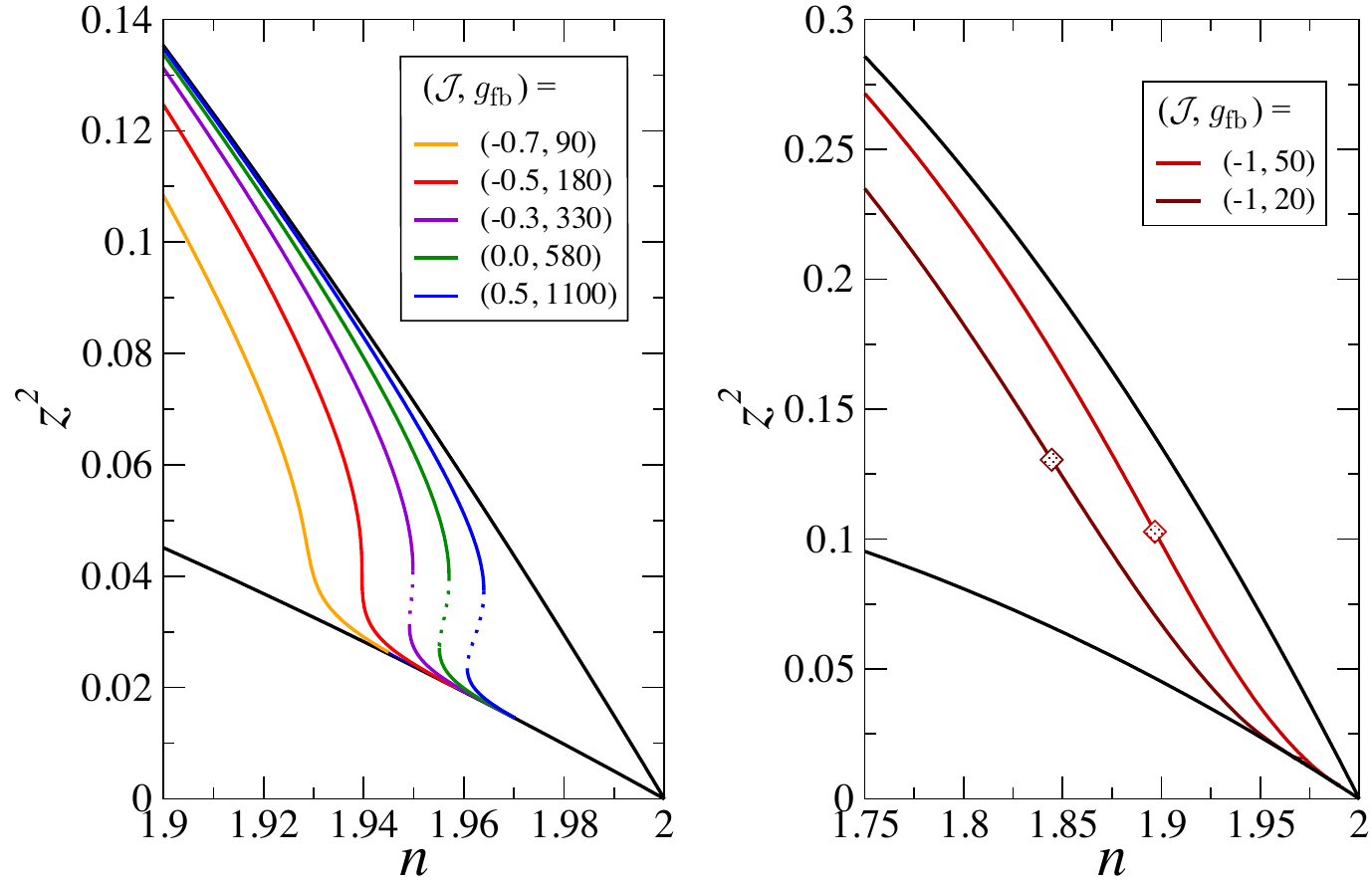}
\caption{Quasiparticle residue from BEG evaluation. The black lines are the upper and lower bounds for $z^2$ from Eq.~(\ref{eq:z2-boundaries}). The two diamonds in the right panel mark the respective inflection points.}
\label{Fig:BEG_z2}
\end{figure}

Most strikingly, $z^2$ displays a transition in this evaluation with BEG
Ising-type fields, the nature of which depends on the strength of nearest
neighbor coupling ${\cal J}$ with respect to $h=J_{\rm H}/2$ (see
Fig.~\ref{Fig:BEG_z2}). We consider all energies in this section in units of
$J_{\rm H}/2$. The coupling parameter $g_{\rm fb}$ mostly shifts the curves
but does not affect the transition qualitatively; its role will be discussed
below.

These results are consistent with those of the slave-boson evaluation (SB) in
the previous section in the sense that we find a continuous as well as a
discontinuous transition in a doping regime close to half-filling. In SB the
type of transition is controlled by the correlation strength $U/W$ (see
Fig.~\ref{Fig:z}). Here the transition is tuned by ${\cal J}$ and ${\cal L}$. 
The dependence of ${\cal J}$, ${\cal K}$ and ${\cal L}$ on the Hubbard-model
parameters $U$, $t,t'$ and $J_{\rm H}$ is rather complex,  and we only
estimate the relative size of the BEG-parameters in
Appendix~\ref{app_BEG_parameters}.

Before we suggest an interpretation of the filling dependence of $z^2$ we
briefly discuss the coupling parameter $g_{\rm fb}$. As said this parameter
shifts the inflection point or the jump in $z^2$: the lower the value of
$g_{\rm fb}$ the farther away the inflection point from half-filling (this is
exemplified in the right panel of Fig.~\ref{Fig:BEG_z2}). In few of the SB results
it appears that $g_{\rm fb}$ is inverse to $U$. This is not unreasonable as a
larger $g_{\rm fb}$, that is, a higher energy cut-off $k_{\rm B} T_{\rm eff}$
accounts for stronger fluctuations in the pseudospin field. Conversely, one
expects that for larger $U$ the slave boson fields are more tightly bound to
the fermionic degrees of freedom and fluctuations of the fields are
suppressed. We introduced $g_{\rm fb}$ phenomenologically and we just use it
to shift the transition structure of $z^2$ to a position compatible with the
SB result.  

The values of $g_{\rm fb}$ in Fig.~\ref{Fig:BEG_z2} are surprisingly large. A brief analysis relates these large values  to the smallness of $z^4$.
To understand this argument, we reparametrize the fermion-boson coupling $g_{\rm fb}$ in terms of a temperature $T_0$ and a $z_0^2= z^2(m_0,q_0)$:
\begin{equation}k_{\rm B} T_{\rm eff}(m,q) = g_{\rm fb}\, z^{4}(m,q)= \frac{k_{\rm B} T_0}{z_0^4}\, z^{4}(m,q)
 \label{eq:T-profile}
\end{equation}
whereby $q_0$ is a reference value which we will choose appropriately and $m_0$ is calculated from the mean-field equation~(\ref{eq:MF-h}) with $m$, $q$ and $T$ replaced by $m_0$, $q_0$ and $T_0$. We choose the reference value $q_0$ such that $q_0 = n_0- 1-  2d_{\rm H}^2$ holds (see Eq.~(\ref{eq:q-dH}))  where we can neglect the small contribution of $d_{\rm H}^2$ for an approximate specification of  $q_0$ in the regime close to half-filling. Now with given $q_0$ and the requirement that the inflection point or jump of $z^2$ is placed in the range of fillings consistent with SB results one identifies values of $k_{\rm B} T_0$  in the range of $0.2-0.4$ and through the relation (\ref{eq:T-profile}) one finds $g_{\rm fb}$ in the range of $20-1100$. The smallness of $z_0^4$  requires large values of $g_{\rm fb}$ in order to fulfill $g_{\rm fb}= (k_{\rm B} T_0/z_0^4)$. 

Qualitatively, the $z^2$-curves of Fig.~\ref{Fig:BEG_z2} resemble those of the SB result in Fig.~\ref{Fig:z}. One might object that $z^2$  calculated within SB theory is notably larger in the metallic regime, especially for $U<W$. This discrepancy, however, is not a consequence of the BEG Ising-type evaluation but it is caused mainly by the neglect of triple occupancies. In fact, the bounds of Eq.~(\ref{eq:z2-boundaries}) (see the black curves in Fig.~\ref{Fig:BEG_z2}) also hold for the SB evaluation if triple occupancies and empty sites are excluded. These neglected contributions are sizable for small and intermediate values of $U$ whereas we are targeting the regime of larger values of $U$ in the BEG scheme. 

The down bending or the jump of $z^2$ to low values close to half-filling is caused by the strong increase of the ``BEG-magnetization'' $m$ in a regime where $q$ approaches one. The upper bound in Eq.~(\ref{eq:z2-boundaries}) stands for $m=0$ whereas the lower bound is determined by full polarization $m=q$. Correspondingly, in the language of the two-band Hubbard model, we observe a transition from a state with smaller orbital polarization ($d_{\rm H}^2 < d_{\rm A}^2 < d_{\rm P}^2$) to a state with strong orbital polarization close to half-filling: $d_{\rm H}^2\simeq 0 \simeq d_{\rm A}^2$ and $2d_{\rm P}^2\gtrsim 1-\delta $ (see Fig.~\ref{Fig:BosOcc} for the BEG result of the filling dependence of  $2d_{\rm P,A,H}^2$ ). Again, the filling dependence of these occupations is qualitatively similar to what was found in the SB evaluation. That inspires the following interpretation of the result of the two-band Hubbard model:

\begin{figure}[!t]
\includegraphics[scale=0.32]{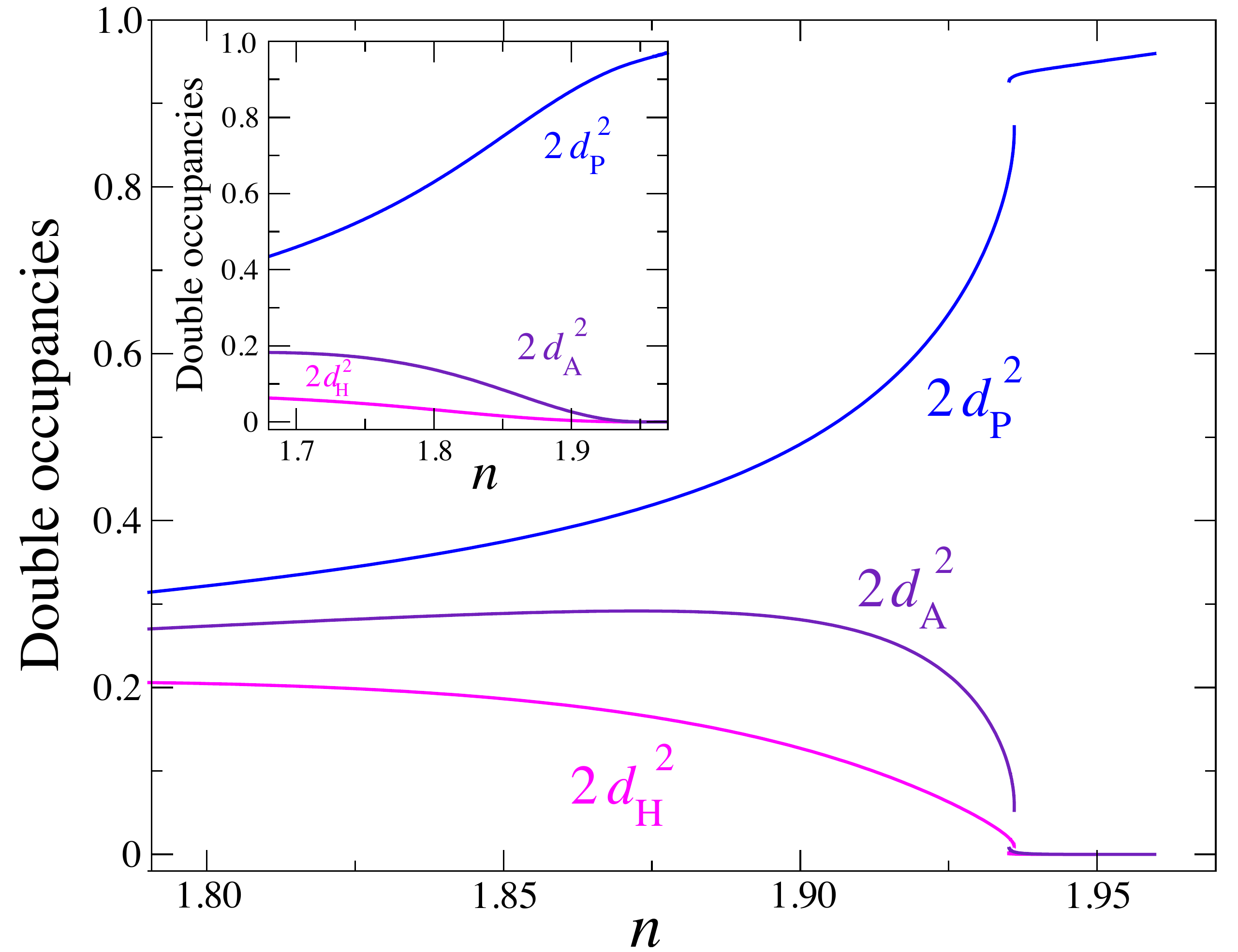}
\caption{Bosonic occupations $2d_{\rm P}^2$, $2d_{\rm A}^2$, and $2d_{\rm H}^2$ in dependence on filling $n$.  The BEG parameters are $h\! = \!1.0$, ${\cal J} \! = \! -0.3$. and  $g_{\rm fb}\! = \!330$.
Inset: $h\! = \!1.0$, ${\cal J}\! = \! -1.0$, and $g_{\rm fb}\! = \!20$.}
\label{Fig:BosOcc}
\end{figure}

Obviously, a finite Hund's  coupling favors a double occupation of sites where
the spins of the two orbitals are aligned ($d_{\rm P}$-state). For strong
coupling bosonic fluctuations to $d_{\rm A,H}$ states are reduced---this is
expressed here through a stronger fermion-boson coupling, that is, through a
smaller $ g_{\rm fb}$ which entails a smaller effective temperature for the
fluctuations in our pseudospin evaluation. Then, the pseudospin magnetization
$m$ is larger. Correspondingly $2d_{\rm P}^2$ is larger and $2d_{\rm A,H}^2$
are smaller for stronger electronic correlations. However
there is a further impact of strong coupling: an orbital (antiferromagnetic)
nearest neighbor coupling  ${\cal J}<0$ becomes effective which induces local
fluctuations to states with antiparallel spins on the two orbitals of a
site. These fluctuations prevent a sharp transition to an orbitally polarized
state: we only observe an inflection point in $z(n)^2$.

For intermediate coupling $ g_{\rm fb}$ is larger and, correspondingly,  the transition is closer to half-filling. Moreover, the reduced antiferromagnetic (orbital) coupling ${\cal J}$ allows Hund's  coupling to dominate in this regime and we identify a discontinuous transition in $z(n)^2$.

In SB theory all single-site double occupancies are represented by bosons, the fluctuations of which are effectively the incoherent background to the (fermionic) quasiparticle excitations.   It depends on the interplay of the fermions and the incoherent (bosonic) background if the reduction of $z^2$ is continuous or discontinuous.

\subsection{Compressibility}
\label{subsec:BEGcomp}

The inverse compressibility is identified from the sum of the inverse compressibilities  of the subsystems whereby each subsystem is characterized by its respective free energy (see, for example, Ref.~\cite{Kopp09}). Each of the free energy terms yields an additive contribution to the inverse compressibility  $\kappa^{-1}$ when forming the second derivative with respect to the total density (and multiplying by a factor density squared). As we keep volume and number of lattice sites $N_{\rm L}$ constant we can use the filling $n$ instead of the density in our evaluation.
There is a fermionic free energy term, which is in fact the fermionic kinetic energy controlled by the inverse effective mass $z^2(n)$, and a pseudospin free energy term originating from the BEG Hamiltonian. 

In an approximation where $d_{\rm H}^2$ is zero we find the simple relation $n= 1+q$ and we can take the derivatives of the pseudospin free energy simply with respect to $q$ to calculate the inverse compressibility. With inclusion of a finite $d_{\rm H}^2$, we have to  respect the relation (\ref{eq:filling}): correspondingly there are corrections from the derivative
\begin{equation}\label{eq:dh-correction}
d n/d q = 1+  2 d(d_{\rm H}^2)/d q
\end{equation}
 that can be  sizable because $d_{\rm H}^2$ decreases rapidly in the doping regime of the continuous transition.

Here the compressibility is to be determined not for given orbital polarization $m$ but for fixed Hund's  coupling, that is, for fixed field $h$. As regards the other BEG-variable, $q$, this is the variable which is related to filling as just discussed. So the appropriate pseudospin free energy depends on $h$ and $q$ which is a Legendre transform of $F( h, \Delta)$ of Eq.~(\ref{eq:BEG-F}) from $\Delta$ to $q$ which we denote as $\Gamma(h, q)$. The derivative of $\Gamma(h, q)/N_{\rm L}$ with respect to $q$ naturally yields $-\Delta(h,q)$, which may be interpreted as the chemical potential related to the $q$-particles. However, as we actually have to take the derivative of $\Gamma(h, q)$ with respect to $n$ and not $q$, we have to multiply the $q$-derivative of $\Gamma(h, q)$  by $ dq/dn$:
\begin{equation}
\frac{d(\Gamma/N_{\rm L})}{dn}= - \Delta\bigl/\bigl(1+2 \frac{d(d_{\rm H}^2)}{d q}\bigr)
 \label{eq:delta-n}
\end{equation}
where all terms depend through $q$ on $n$. The contribution of the pseudospins to the inverse compressibility, $\kappa_{\rm ps}^{-1}$, is now the derivative of this pseudospin chemical potential with respect to $n$:
\begin{equation}
(n^2\kappa_{\rm ps})^{-1} = 
\frac{d^2(\Gamma/N_{\rm L})}{dn^2} = -
 \frac{d}{dn} \biggl(\Delta\bigl/\bigl(1+2 \frac{d(d_{\rm H}^2)}{d q}\bigr)\biggr)
 \label{eq:ps-comp}
\end{equation}
Here $\Delta(h,q)$ is found from the mean-field expression Eq.~(\ref{eq:MF-delta}) with $m$ replaced by its $q$-dependent mean-field value 
$m(q)$, and the temperature is replaced by $T_{\rm eff}$ of Eq.~(\ref{eq:T-profile}) in that mean-field evaluation. It is evident that the $n$-dependence of the term in parentheses on the rhs of Eq.~(\ref{eq:ps-comp}) has to be determined first from Eq.~(\ref{eq:filling}) and the mean-field equations, before the derivative with respect to $n$ can be calculated.

The fermionic contribution to the inverse compressibility,   $\kappa_{\rm qp}^{-1}$, results directly from the second derivative of the kinetic energy $z(n)^2 E_{\rm kin}(n)$ with respect to $n$. In order to have a simple analytical expression for $E_{\rm kin}(n)$  we take the dispersion from Eq.~(\ref{dispersion_0}) with $t'=0$. As the second derivative of the kinetic energy is dominated by the curvature of $z(n)^2$ in the transition regime, the filling dependence of the unrenormalized $E_{\rm kin}(n)$ is of little consequence if it is sufficiently smooth. This is the case, as the dispersion integrates to the smooth function $
E_{\rm kin}(n)/N_{\rm L}= -\frac{2}{\pi} W \cos\bigl[\frac{\pi}{4} (2-n)\bigr]
$ 
(compare Eq.~(\ref{eq:Fkin})
where $W$ is the bandwidth and the two spin directions have been taken care of by a factor 2. The fermionic compressibility is now
\begin{equation}
(n^2\kappa_{\rm qp})^{-1} = 
\frac{d^2(z^2 E_{\rm kin}/N_{\rm L})}{dn^2}
\label{eq:fermi-comp}
\end{equation}
In order to put $\kappa_{\rm qp}^{-1}$ in relation to $\kappa_{\rm ps}^{-1}$ quantitatively, we have to fix $W$: with $J_{\rm H}= W/6$ used in the section on the SB results and $h= J_{\rm H}/2$, we choose correspondingly $W=12 h$ for the further evaluation of the total compressibility.

Eventually, the total compressibility $\kappa_{\rm tot}$ is determined from the subsystem compressibilities Eqs.~(\ref{eq:ps-comp}) and (\ref{eq:fermi-comp}) through
\begin{equation}
\kappa_{\rm tot}^{-1} = \kappa_{\rm ps}^{-1} + \kappa_{\rm qp}^{-1} 
\end{equation}
As is obvious from Eq.~(\ref{eq:ps-comp}) we now need $\Delta$ which is extracted from Eq.~(\ref{eq:MF-delta}) where $T$ has to be replaced by $T_{\rm eff}$. The evaluation of $\Delta$ requires to choose an appropriate $\cal K$-parameter of the BEG-model. As one can learn from Eqs.~(\ref{eq:parameters}) in Appendix~\ref{app_BEG_parameters} and the following discussion, the parameter $\cal K$ is positive and  considerably larger than $|{\cal J}|$. We take  ${\cal K}= 8.0$ (again in units of $J_{\rm H}$). 

\begin{figure}[!t]
\includegraphics[scale=0.30]{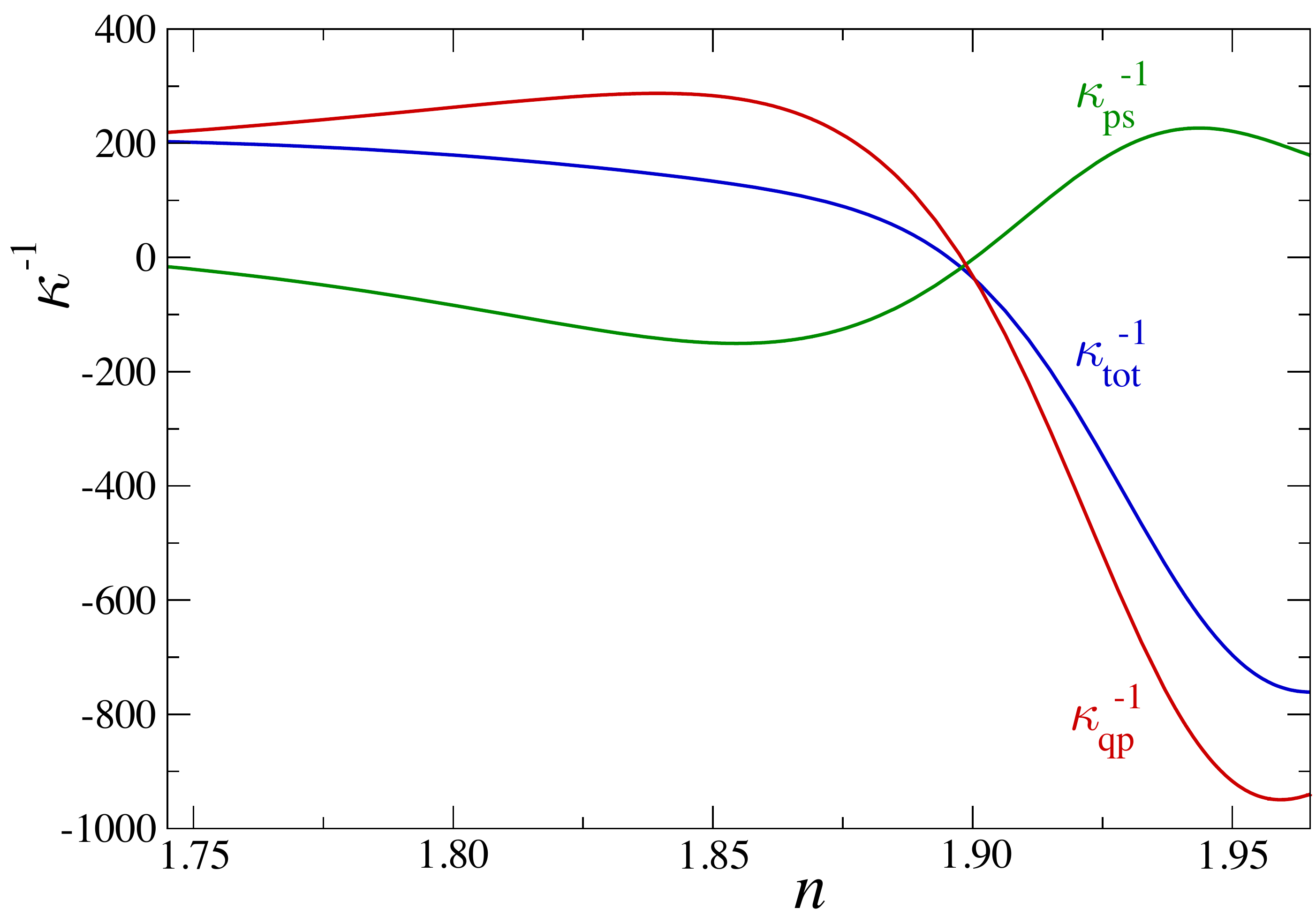}
\caption{Compressibility in dependence on filling $n$.  The BEG parameters are $h\! = \!1.0$, ${\cal J} \! = \! -1.0$, $g_{\rm fb}\! = \!50$ and ${\cal K}=8.0$. Here $\kappa^{-1}$ is in units of $J_{\rm H}/2 a^2$ and the coupling parameters are in units of  $J_{\rm H}/2$.}
\label{Fig:kappa-1p0}
\end{figure}

The compressibilities are displayed in Fig.~\ref{Fig:kappa-1p0} for ${\cal J}= -1.0$. That value of ${\cal J}$ entails that the transition in $z^2(n)$ is continuous. This is now reflected in a continuous transition of the compressibility  $\kappa_{\rm tot}$ from positive to negative values close to half-filling.

\begin{figure}[!t]
\includegraphics[scale=0.29]{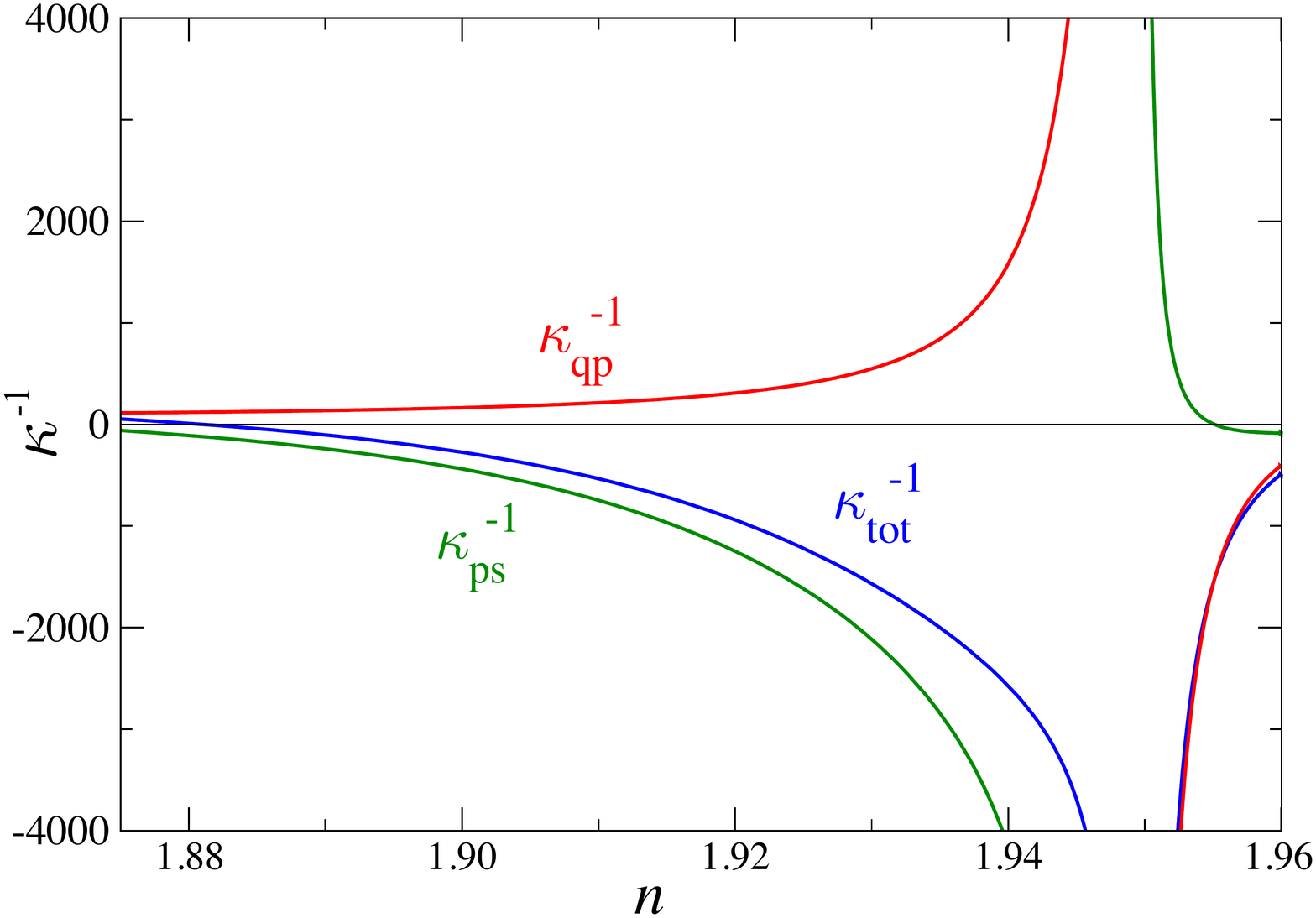}
\caption{Compressibility in dependence on filling $n$.  The BEG parameters are $h\! = \!1.0$, ${\cal J} \! = \! -0.3$, $g_{\rm fb}\! = \!330$ and ${\cal K}=8.0$.
Here $\kappa^{-1}$ is in units of $J_{\rm H}/2 a^2$ and the coupling parameters are in units of  $J_{\rm H}/2$.}
\label{Fig:kappa-0p3}
\end{figure}

Similarly, the compressibilities are discontinuous for ${\cal J}= -0.3$ as seen  in Fig.~\ref{Fig:kappa-0p3}. This is expected as the quasiparticle weight $z^2$ is discontinuous for  these less negative values of ${\cal J}$. 

It is obvious that in the considered filling regime the inverse pseudospin compressibility $\kappa_{\rm ps}^{-1}$ partially cancels the inverse quasiparticle compressibility $\kappa_{\rm qp}^{-1}$. The exact degree of cancellation depends on the relative values of the partial compressibilities.
However in the vicinity of the transition, $\kappa_{\rm qp}^{-1}$ becomes already negative when $\kappa_{\rm ps}^{-1}$ is still negative (see Fig.~\ref{Fig:kappa-1p0}). This behavior is analogous to what was observed in the SB formalism for $\kappa_{\rm f}^{-1}$ and $\kappa_{\rm b}^{-1}$. If the zero crossings of both were at the same filling $n$ then the transition to bad metal behavior (the inflection point of $z^2$) would coincide with the transition to negative compressibility. Instead we find two distinct transitions. 

The negative compressibility of the pseudospin subsystem indicates that it is not in its (thermodynamic) equilibrium which is presumably true also for the bosonic subsystem in SB theory---however, a partition into subsystems is not evident there. For the pseudospin subsystem we find a mean-field solution for given $q$ and the appropriate effective temperature but this solution does not represent the global minimum for $q>0.744$ (assuming here ${\cal J}=-1.0$). The chemical-potential parameter $-\Delta$ of Eq.~(\ref{eq:MF-delta}) is tied to the pseudospin field $q$,  which is fixed by the choice of $n$. Actually  a considerably lower  $q$ represents the thermodynamically stable state for that value of $\Delta$. If $n$ and, correspondingly, $q$ were not fixed then the pseudospin system would relax to this lower $q$. The thermodynamic stability of the BEG subsystem is discussed in Appendix~\ref{app_BEG_phase_diagram}. It is the requirement of sufficiently large $q$ 
to keep a filling in the low doping regime and of sufficiently small effective temperature $T_{\rm eff}$ enforced by small values of $z^2$ close to half filling that drives the pseudospin out of equilibrium: there is no global minimum of the thermodynamic potential in this regime.

Close to half-filling  $\kappa_{\rm ps}^{-1}$ becomes positive for ${\cal J}=-1.0$ (see Fig.~\ref{Fig:kappa-1p0}). This behavior close to the metallic transition is triggered by the strong filling dependence of the fields that represent double occupancies.
In particular the field $d_{\rm H}^2$ enters through the factor $dq/dn$ of Eq.~(\ref{eq:dh-correction}) into Eq.~(\ref{eq:delta-n}) and reverses the slope of the pseudospin chemical potential term $d(\Gamma/N_{\rm L})/dn$ with respect to $n$ in the bad metal state. This signifies that $\kappa_{\rm ps}^{-1}$ is positive there. However, the pseudospin subsystem is still not in its (thermodynamic) equilibrium.

For the discontinuous case  (see Fig.~\ref{Fig:kappa-0p3}) $\kappa_{\rm ps}^{-1}$
becomes again negative in the vicinity of half-filling as opposed
to the behavior of $\kappa_{\rm b}^{-1}$ in the SB evaluation. In this respect
we reemphasize that the BEG-results cannot be trusted for lower values of
$U/W$. In fact,  effective interactions between doubly occupied sites were
parameterized as nearest neighbor pseudospin exchange within a perturbative
scheme where triple and quadruple occupancies are suppressed. 

On the other side, for $U/W\gg 1$, the BEG parameters ${\cal J}$, ${\cal K}$, and ${\cal L}$  become small  in comparison to the fixed $h=J_{\rm H}/2$ (see Appendix~\ref{app_BEG_parameters}) and one might expect a reemergence of the first order transition. However,
for large values of $U/W$, the chosen ``temperature profile'' with exponent $\alpha =2$ may have to be  modified to $\alpha =1$, as discussed at the end of Sec.~\ref{subsec:BEGmodel}. In that case, we find exclusively a continuous transition---in line with the strong coupling SB results.

Correspondingly, we expect a parameter window for intermediate to moderately strong
correlation strength where the results from this phenomenological approach
are qualitatively valid. 

Overall the results from the BEG modeling can be compared reasonably well with those of the SB theory. They explain the transition in $z^2(n)$ through a transition in the pseudospin magnetization controlled by $h$, that is, in the orbital polarization of the 2-band Hubbard model 
controlled by $J_{\rm H}$. Also the transition to the negative
compressibility in the low doping regime is  recovered. However, it is not a phase transition of the pseudospin system of the BEG model but it is the  feedback mechanism between the pseudospin system and the (fermionic) quasiparticle system which causes the transitional behavior.

\section{Conclusions and Outlook}
\label{sec:conclusion}

This work is concerned with the two-band Hubbard model in the presence of a
finite Hund's coupling $J_{\rm H}$. In particular, we investigated the
paramagnetic state  close to half filling ($n=2$) using an extended
Kotliar-Ruckenstein slave-boson technique. Previously, a first order
transition with a coexistence regime between a metallic and a bad metal state
below a critical point  was identified~\cite{Fresard2001} and, more
recently,  a continuous transition signaling a charge instability for larger
on-site interaction $U$ was discovered~\cite{Medici17}. Both transitions were
considered in dependence on $J_{\rm H}/U$ and it was found that they are
absent for $J_{\rm H}=0$. 

Here, we analyzed these transitions jointly for fixed $J_{\rm H}$ and
found that the line related to the continuous transition, characterized by
zero inverse compressibility, merges with the first order transition at a
critical end point (CEP). This CEP is close to the CP. Beyond the CEP, that is
for the filling range towards half filling, the charge instability persists,
however only in the metallic state which is not the global free energy minimum
in this range of $n$. 

The inverse compressibility jumps from positive
to negative values jointly with the inverse effective mass along the first
order transition line. This transition into the negative-compressibility
bad-metal regime extends to smaller values of $U$ down to the metal-insulator
transition at half filling where it ends at the Mott-insulator transition for
the two-band Hubbard model.

A  recent DMFT-based work~\cite{Chat22} on the two-orbital Hubbard model
that presents  the phase diagram close to half filling in dependence on $J_{\rm H}/U$
compares well with the slave boson findings of ~\cite{Fresard2001} concerning the first order
transition with a coexistence regime and a (quantum) critical point. In the regime where Ref.~\cite{Chat22} 
has ``no solution'' we identify a critical end point (CEP). The QCP in their work is the CP in our work but our high resolution in the parameter space allows to separate the charge instability line from the CP in our Fig.~\ref{Fig:pd_zoomed} whereas in \cite{Chat22}  the QCP is directly connected to the ``crossover (enhanced compressibility)'' line in Fig.~2a of Ref.~\cite{Chat22}.
More refined evaluations may help to decide which scenario is realized in these systems.

The slave boson theory suits well to distinguish between the excitations into
coherent (fermionic) quasiparticles and multiparticle or collective (bosonic)
excitations. Even though this is implemented here only on the saddle point
level, the decomposition allows in this context to study the quasiparticle
contribution to the compressibility separately from the bosonic background as
the inverse compressibility can be split into the corresponding partial
(inverse) compressibilities. The quasiparticle compressibility is controlled
by the curvature and jump of the inverse effective mass $\propto z^2(n)$. In
contrast, the bosonic contribution in this regime is governed by the bosonic
field which represents doubly occupied sites with parallel spins in the two
distinct orbitals, that is, $d^2_{\rm P} (n) $. The Hund's coupling $J_{\rm
  H}$ triggers both, the sharp drop in $z^2(n)$ and the steep increase in the
double occupancy $d^2_{\rm P} (n) $ when approaching half filling. This drop
and increase for $z^2(n)$ and $d^2_{\rm P} (n)$, respectively, may be realized
by a jump or an inflection point in their $n$-dependence. Obviously, Hund's
coupling favors this type of double occupancy ($d^2_{\rm P} $ ) energetically
thereby suppressing not only the competing double occupancy configurations but
also the triple occupancy and concomitantly the single occupancy so
effectively that a phase transition is accomplished---either first order or
continuous. 

We confirmed that in the absence of Hund’s coupling, the curvature of
$z^2(n)$ close to half filling does not switch its sign, i.e., $z^2(n)$
stays concave. The filling range where $z^2(n)$ becomes convex, which
signals a charge instability and negative compressibility, is in fact
controlled by the size of $J_{\rm H}$·

In view of the applied technique and our focus on low doping and orbital ordering in
this regime we may now substantiate the assumption that the Zeeman-like term to Hund’s coupling is the dominant contribution in the saddle point approximation. It generates a transition into the phase with parallel spin orientation of the electrons in the two orbitals close to half filling. The spin-flip term that is also present in Hund’s coupling introduces fluctuations into the local configuration with antiparallel spins in the two orbitals. It favors an antiparallel configuration which, however, is only sustained in $J_{\rm H}^2/U$. As we considered only small $J_{\rm H}$ with respect to $U$, this correction is of minor relevance:  it may reduce the regime of the negative compressibility state slightly~\cite{Medici17}. The third term in Hund’s coupling, the pair-hopping term, favors the configuration with double occupancy on a single orbital. However the energy of this state is already considerably higher in energy (by $3 J_{\rm H}$ in a local estimate) and is therefore disfavored. Moreover a constraint (see Eq.~(\ref{eq:relds})) enforces this configuration to vanish if the configuration with antiparallel spins is suppressed close to half filling.

In that respect our approach is consistent as  $J_{\rm H}/U$  is still small in the relevant part of the phase diagram (see Fig.~\ref{Fig:pd}).
Upon increasing $J_{\rm H}$ from $W/6$ to $W/3$ moves the charge instability line CIL (green line in Fig.~\ref{Fig:inst}) down towards lower values of $U$ (about half its displayed value) in the regime of the rather horizontal extension of the CIL but also bends the vertical part of the curve towards lower doping values at large $U$. The maximal doping value of the CIL is increased to approximately $0.18$. This may still be seen as a correction to the displayed results. However, for $J_{\rm H}$ as large as $W/2$, the regime of charge instability would already form for $U$ considerably less than $W$ and results with the approximate Zeeman-type Hund’s coupling would not be trustworthy.

Breaking particle-hole symmetry does not change the results qualitatively. 
The origin of the phase transitions is to be related to Hund’s coupling and its impact on the quasiparticle residue $z^2(n)$ which becomes small and convex close to half-filling. Particle-hole symmetry has no particular significance in that respect. Introducing a van Hove singularity close to half-filling on either side of the center of a band suppresses the inverse compressibility related to the kinetic term for filling in this regime. It may be worthwhile to investigate such a situation.

Non-local correlations may arise from further non-local terms in the Hamiltonian or from contributions beyond the 
slave-boson saddle-point approximation. For the one-band model the additional contribution of a non-local exchange (Fock) term to the compressibility was investigated~\cite{Steffen17}. As expected, the compressibility is strongly affected in the low-density case but beyond that it is well presented by the local mean-field terms. Given that the explicit inclusion of a non-local term does not drive the saddle-point physics 
for intermediate densities into a different regime we conjecture that non-local correlations do not present substantial corrections
in the compressibility with or without non-local Hamiltonian terms. In particular, this will apply for our modelling close to half filling where the transition to negative compressibility is controlled by the filling dependence of $z^2(n)$; fluctuations of $z^2(n)$ through non-local correlations are not pivotal as long as one does not consider the regime in the immediate vicinity of half filling.  

In a toy-model approach, the most prominent bosonic degrees of freedom were
mapped onto Ising-like (spin-1) pseudospins within a Blume-Emery-Griffiths
model that implements the Hund's coupling by a Zeeman-like field and the
correlations through quadratic and bi-quadratic nearest-neighbor
exchange-energy terms. This allows to discuss these degrees of freedom in a
classical model although the Ising fields are then coupled to the (fully
quantum mechanical) quasiparticle system through a feedback mechanism. It
appears in this approximate treatment that the transition is not a phase
transition of the pseudospin system itself but it is a transition generated by
the feedback of the quasiparticles and vice versa. The pseudospin system by
itself is rather out of equilibrium close to half filling and the coupling to
the fermionic system keeps it in this state. It is only the joint
pseudospin-quasiparticle system that, for fixed filling $n$, is in
equilibrium. 

The capacitance of a heterostructure device is intimately related to
electronic compressibilities of the electrodes. As the
compressibility of this two-band Hubbard-model electronic system depends very
sensitively on the electron density close to half filling---also under the
consideration of the transitions to negative compressibility---it is well
conceivable that  micro-device capacitances may be very effectively controlled
and switched through small electronic-density variations. 

We also suggest the intriguing possibility to switch between low and high
capacitance through electric pulse switching between the high resistance Mott
insulator and the low resistance metallic state~\cite{Cario10}.

\begin{acknowledgments}
Financial support by the Deutsche Forschungsgemeinschaft (project number
107745057, TRR 80) is gratefully acknowledged. R.F. is grateful for
the warm hospitality at the University of Augsburg where part of this work
has been done, and to the R\'egion Normandie for financial support. 

\end{acknowledgments}

\appendix

\section{Further slave boson properties}
\label{app_GS}

The generic slave boson rewriting of the physical electron
creation operators in terms of auxiliary particles Eq.~(\ref{Eq:forc}) makes it
manifest that any slave boson representation possesses an internal gauge
symmetry group~\cite{RN83a,RN83b,NR87,Fre01,Kop07,Kop12,fresard12,Dao20}. 
In the present case of the two-band model and using the above four-valued
spin-band index $\alpha$ the representation of the physical electron operators
Eq.~(\ref{Eq:forc}) is invariant under the gauge transformations 
\begin{equation}
  \left\{ \begin{array}{l}
          f_\alpha \longrightarrow \ee^{-\ii \chi_\alpha} 
          f_\alpha
          \\
          e_{\phantom{\alpha}} \longrightarrow \ee^{\ii \theta} e
          \\
          p_{\alpha} \longrightarrow \ee^{\ii \left(\chi_\alpha + \theta \right)}
          p_{\alpha}
          \\
      d_{\alpha,\alpha'} \longrightarrow \ee^{\ii \left(\chi_{\alpha} + 
      \chi_{\alpha'}  + \theta\right)} d_{\alpha,\alpha'}\\
      t_{\alpha,\alpha',\alpha''} \longrightarrow \ee^{\ii \left(\chi_{\alpha} + 
    \chi_{\alpha'} + \chi_{\alpha''}  + \theta\right)} t_{\alpha,\alpha',\alpha''}
          \\
          \varpi_{\phantom{\alpha}} \longrightarrow \ee^{\ii (\theta + \sum_{\alpha}\chi_{\alpha}) } \varpi
         \end{array}\,.
   \right.      
\end{equation}
The gauge symmetry group is therefore $U(1) \times U(1) \times U(1) \times
U(1) \times U(1)$.
The Lagrangian also possesses this symmetry. 
Expressing the bosonic fields in amplitude and phase variables as
\begin{align}
e(\tau) &=  \sqrt{R_e(\tau)}\, \ee^{\ii \theta(\tau)}\nonumber\\
p^{\phantom{\dagger}}_\alpha (\tau) &=  \sqrt{R_\alpha(\tau)}\,
\ee^{\ii (\chi_\alpha(\tau) + \theta(\tau))} 
\end{align}
allows to gauge away the phases of the above five slave boson fields, 
provided one introduces the five time-dependent Lagrange multipliers
\begin{align}
\lambda' (\tau) &\equiv  \lambda' + \partial_{\tau}
\theta(\tau) \nonumber\\
\lambda_\alpha (\tau) &\equiv  \lambda_{\alpha}
  - \partial_{\tau} \chi_{\alpha}(\tau) .
\end{align}
Here the radial slave boson fields are implemented in the continuum
limit~\cite{RN83a,RN83b,NR87}, but introducing radial slave boson fields can
also be achieved in the discrete time step set-up~\cite{Kop07,Kop12,Dao20}.

As these bosonic fields have been deprived of their phase degree of freedom
they do not undergo Bose condensation any longer. In fact, their exact
expectation values are generically non-vanishing~\cite{Kop12} (see
Ref.~\cite{Kop07} in the case of Barnes' representation to the single impurity Anderson model), and may be approximately obtained through
the saddle-point approximation (SPA) that we used above. 
This approximation is exact in the
large degeneracy limit, with Gaussian fluctuations generating the $1/N$
corrections~\cite{Fresard1992} (for a recent detailed reference, see
Ref.~\cite{Dao17}). It has been tested against quantum Monte Carlo 
simulations in the most challenging $N=2$ case: A quantitative agreement for
charge structure factors was demonstrated \cite{Zim97} and, for example, a
very good agreement on the location of the metal-to-insulator transition for
the honeycomb lattice has been shown \cite{Doll3}. Also the comparison of
ground state energies to numerical simulations are
excellent~\cite{Fre91}. Further quantitative 
agreement of ground state energies and site-dependent local magnetization with 
density matrix embedded theory have been recently reported~\cite{Riegler20}. 

We now turn to the operator $z_{i,\alpha}$. It represents the change in the bosonic occupations
which results from the annihilation of an electron (see of Eq.~(\ref{Eq:forc})).
 In the considered paramagnetic phase one  introduces  $z$ 
through $z \equiv z_{i,\alpha}$. Following Ref.~\cite{Fresard1997} it reads:
\begin{equation}\label{Eq:defz}
z=L \tilde{z} R
\end{equation}
with
\begin{subequations}
\label{Eq:defsLR}
\begin{align}
\tilde{z}&=e p + \left( p + t \right)\left( d_{\rm P} + d_{\rm A} + d_{\rm H} \right) + t \varpi\\
L&=\left(1 - p^2- d_{\rm P}^2 - d_{\rm A}^2 - d_{\rm H}^2 - 3 t^2 - \varpi^2
\right)^{-\frac12} = \left(1 - n_{\alpha}\right)^{-\frac12}\\
R&=\left(1 - e ^2 - 3 p^2- d_{\rm P}^2 - d_{\rm A}^2 - d_{\rm H}^2 - t^2 \right)^{-\frac12}= n_{\alpha}^{-\frac12}
\end{align}
\end{subequations}
and $n_{\alpha} \equiv \frac{n}{4}$. Note that $z$ depends on the three $d$-fields in a symmetric fashion.

\section{Saddle-point equations}
\label{app_SP}

The saddle-point equations associated to the
derivative with respect to the bosons read:
\begin{widetext}
\begin{align}
\lambda'+ \big[ p + \frac{\tilde{z} e}{n_{\alpha}}\big]
\frac{2\mathcal{B}}{e}&= 0 \label{eq:spae} \\
\lambda' - \lambda + \big[ e + d_{\rm P} + d_{\rm A} + d_{\rm H}
+ \tilde{z} p \frac{3-2n_{\alpha}}{n_{\alpha}(1 - n_{\alpha})}\big]\frac{\mathcal{B}}{2p}
&= 0
 \label{eq:spap}\\
\lambda' - 2\lambda + U_{\rm P} + \big[ p+t + 
\frac{\tilde{z}d_{\rm P}}{n_{\alpha}(1 - n_{\alpha})}\big]\frac{\mathcal{B}}{d_{\rm P}}
&= 0
 \label{eq:spadp}\\
\lambda' - 2\lambda + U_{\rm A} + \big[ p+t + 
\frac{\tilde{z}d_{\rm A}}{n_{\alpha}(1 - n_{\alpha})}\big]\frac{\mathcal{B}}{d_{\rm A}}
&= 0 \label{eq:spada}\\
\lambda' - 2\lambda + U_{\rm H} + \big[ p+t + 
\frac{\tilde{z}d_{\rm H}}{n_{\alpha}(1 - n_{\alpha})}\big]\frac{\mathcal{B}}{d_{\rm H}}
&= 0
 \label{eq:spadh}\\
\lambda' - 3\lambda + U_{\rm P}+ U_{\rm A} + U_{\rm H} + 
\big[ \varpi + d_{\rm P} + d_{\rm A} + d_{\rm H} +
\frac{\tilde{z}t(1 + 2 n_{\alpha})}{n_{\alpha}(1 - n_{\alpha})}\big]\frac{\mathcal{B}}{2t}
&= 0
 \label{eq:spat}\\
\lambda' - 4\lambda + 2 \big(U_{\rm P}+ U_{\rm A} + U_{\rm H}\big) + 
\big[ t + \frac{\tilde{z}\varpi}{(1 - n_{\alpha})}\big]\frac{2\mathcal{B}}{\varpi}
&= 0
 \label{eq:spaq}
\end{align}
\end{widetext}
where we introduced
\begin{align}
\bar{\epsilon} &\equiv \sum_{{\bf k},\nu} f_F(E_{{\bf k},\nu})\epsilon^{(0)}_{{\bf k},\nu}
\nonumber\\
\mathcal{B} &\equiv \frac{\tilde{z}\bar{\epsilon}}{n_{\alpha}(1 - n_{\alpha})}\,.
\end{align}
Here, $f_F(\ldots)$ is the Fermi function. Steps towards the solution of the
saddle-point equations involve solving Eqs.~(\ref{eq:spae}, \ref{eq:spaq}) with
respect to $\lambda$ and $\lambda'$. One finds:
\begin{align}
\lambda &=\frac{U_{\rm P}+ U_{\rm A} + U_{\rm H}}{2} + \Big(\frac{t}{\varpi} -
\frac{p}{e} + 
\frac{\tilde{z}(2n_{\alpha}-1)}{n_{\alpha}(1 - n_{\alpha})}\Big) 
\frac{\mathcal{B}}{2}\nonumber\\
\lambda' &= - 2\Big(\frac{p}{e} + \frac{\tilde{z}}{n_{\alpha}}\Big) \mathcal{B}
\label{eq:sollam}
\end{align} 
Inserting these solutions into
Eqs.~(\ref{eq:spadp}, \ref{eq:spada}, \ref{eq:spadh}) allows to write :
\begin{align}
U_{\rm H} - U_{\rm P} &= \Big(\frac{1}{d_{\rm P}} - \frac{1}{d_{\rm H}}\Big) 
(p+t)\mathcal{B}\nonumber\\
U_{\rm A} - U_{\rm P} &= \Big(\frac{1}{d_{\rm P}} - \frac{1}{d_{\rm A}}\Big)
(p+t)\mathcal{B}\nonumber\\
U_{\rm H} - U_{\rm A} &= \Big(\frac{1}{d_{\rm A}} - \frac{1}{d_{\rm H}}\Big)
(p+t)\mathcal{B}\,.
\end{align} 
A useful relation between the three $d$-bosons may be derived out of these
equations: 
\begin{equation}\label{eq:relds}
d_{\rm H} = \frac{d_{\rm P}d_{\rm A}}{3d_{\rm P}-2d_{\rm A}}
\end{equation}
which both eases the numerical task and the interpretation of the results.
Further steps towards the solution of the saddle-point equations arise from
the derivatives with respect to the Lagrange multipliers. They read:
\begin{align}
e^2 + 4 p^2 + 2 \left(d^2_{\rm P} + d^2_{\rm A} + d^2_{\rm H}\right) +
4t^2+\varpi^2-1&=0 \\
4 p^2 + 4 \left(d^2_{\rm P} + d^2_{\rm A} + d^2_{\rm H}\right) +
12t^2+4\varpi^2 -n &=0
\end{align}
They may be solved with respect to $e$ and $\varpi$ as:
\begin{align}
e^2&=1 - n_{\alpha}-\big(3 p^2 + d^2_{\rm P} + d^2_{\rm A} + d^2_{\rm H}+ t^2 \big) \label{eq:sole}\\
\varpi^2&=n_{\alpha}-\big(p^2 + d^2_{\rm P} + d^2_{\rm A} + d^2_{\rm H}+ 3t^2 \big)\,.\label{eq:solq}
\end{align}
Altogether, one is left with four unknowns ($p$, $d_{\rm P}$, $d_{\rm A}$, and
$t$) determined by Eqs.~(\ref{eq:spap}, \ref{eq:spadp}, \ref{eq:spada},
\ref{eq:spat}) rewritten using Eqs.~(\ref{eq:sollam}, \ref{eq:sole}, 
\ref{eq:solq}).  

\begin{figure}[t]
\includegraphics[scale=0.42]{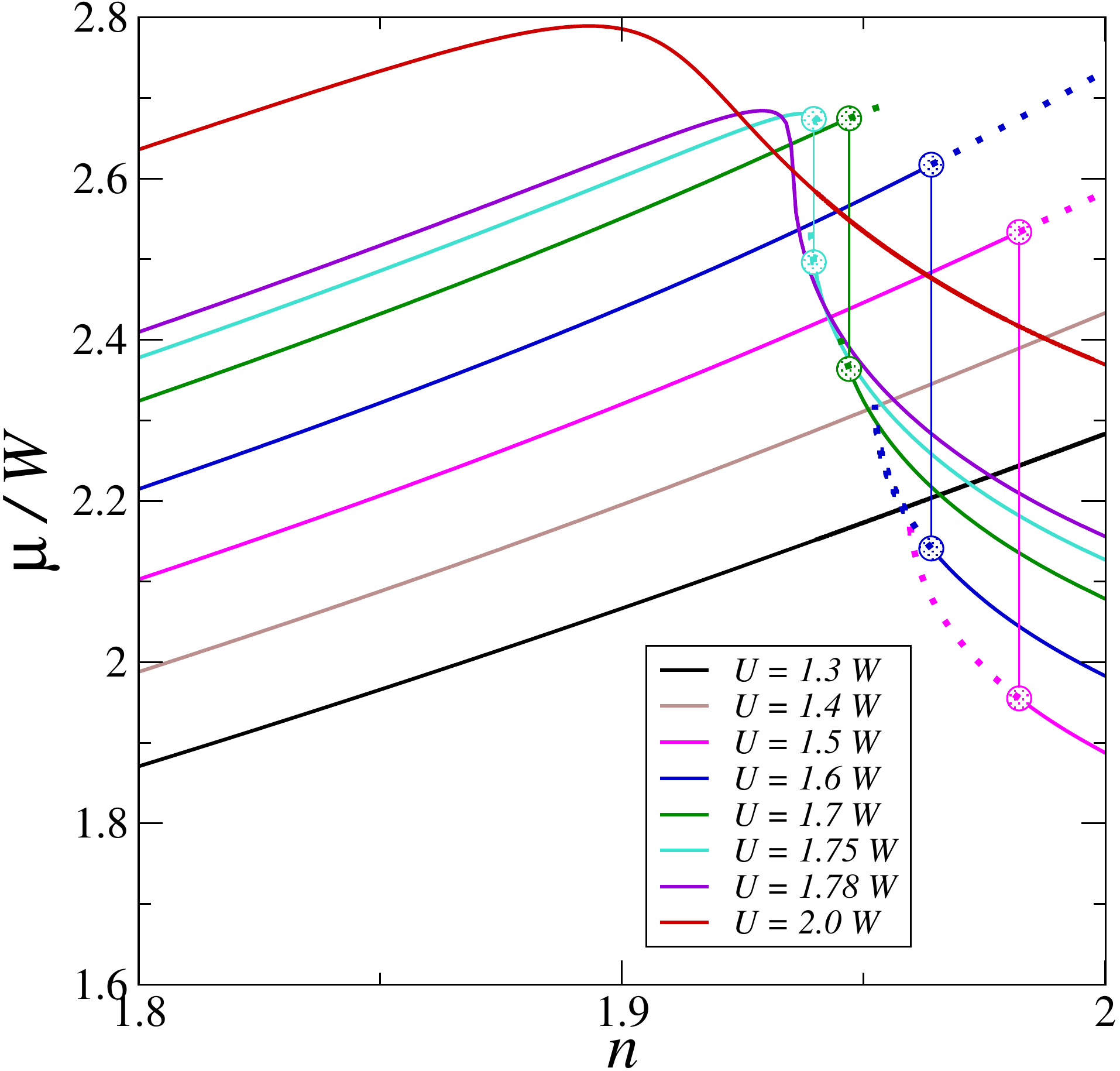}
\caption{
Chemical potential in dependence on filling $n$ for $J_{\rm H} = W/6$. The circles and the
  vertical thin lines mark the first order transitions. The dotted curves refer to 
  metastable states.
  }
\label{Fig:mu}
\end{figure}

\section{Chemical potential}
\label{app_mu}

A key quantity that reveals the addressed first order
transition is the chemical potential which is depicted in
Fig.~\ref{Fig:mu}. For $U<U_{\rm MI}\simeq 1.41~W$ and starting from a large hole doping value,
it is found that the chemical potential monotonically grows with increasing
density, which ensures positive electronic compressibility and
thermodynamical stability of this coherent metallic phase,
as indicated by its quasiparticle residue $z^2>0.5$. Furthermore $\mu$
monotonically increases with $U$. 
However, if $U$ exceeds $U_{\rm MI}$, a bad metal state stabilizes close
to half filling and $\mu$ jumps to a lower value and decreases further towards
half filling. 

If $U$ exceeds $U_{\rm CEP}$  the chemical potential first grows but then reaches
a maximum and  decreases until the metallic solution ceases to
exist. Accordingly, the density dependence of $\mu$ reveals a charge
instability---signaled by the resulting negative electronic
compressibility. This metallic, negative compressibility state is superseded 
in a first order transition to a bad metal state when $n$ is further increased.  

Above $U_{c}^*$ the continuity of the density
dependence of the chemical potential is restored. Though continuous, these
curves are characterized by a maximum which implies that the charge instability persists
above $U_{c}^*$. The latter is a hallmark of the doped Mott insulator.

\section{ Single and triple occupancies}
\label{app_st}

\begin{figure*}[t]
\subfloat[]{
  \includegraphics[scale=0.50]{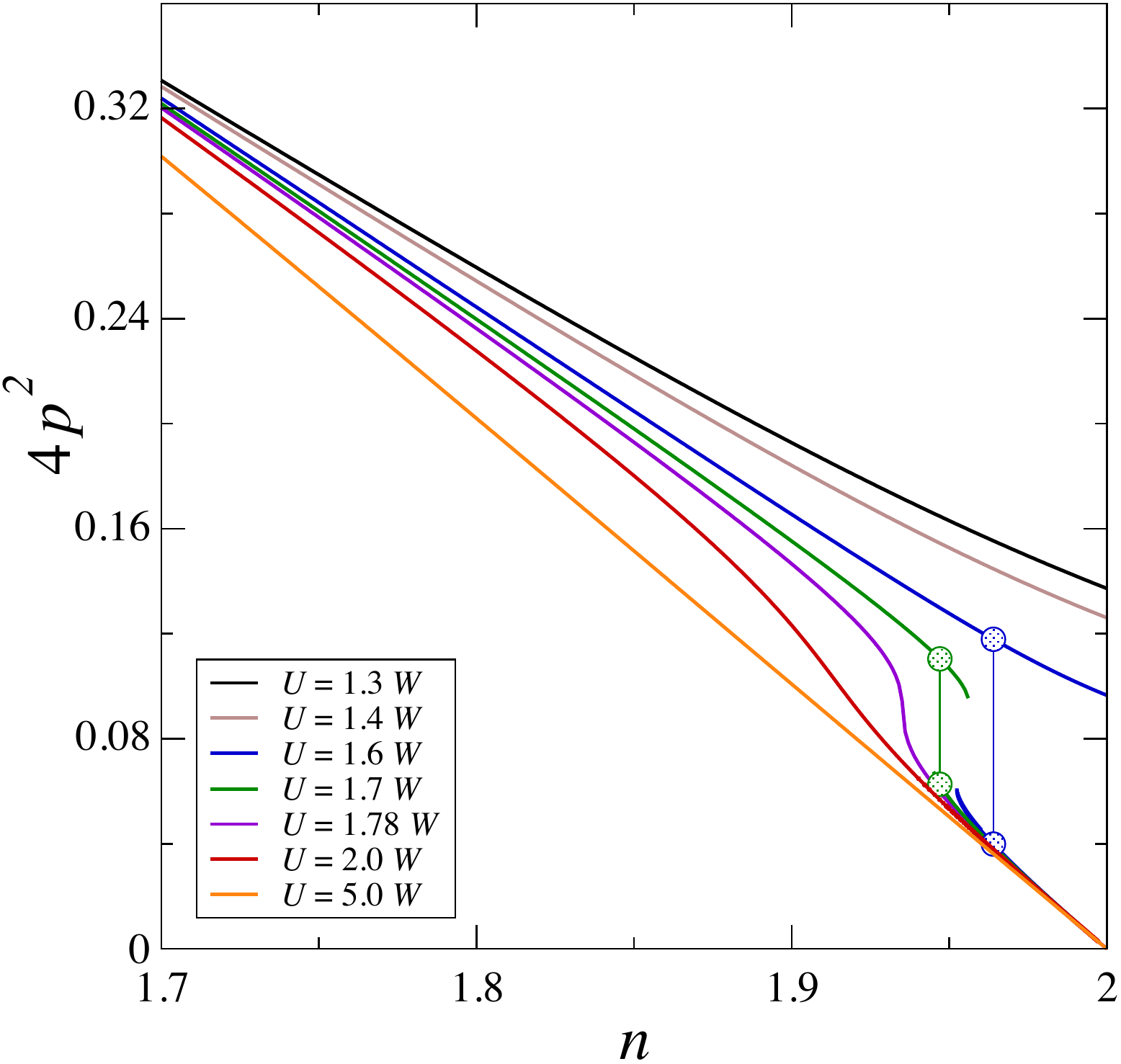}
}\hfil
\subfloat[]{
  \includegraphics[scale=0.50]{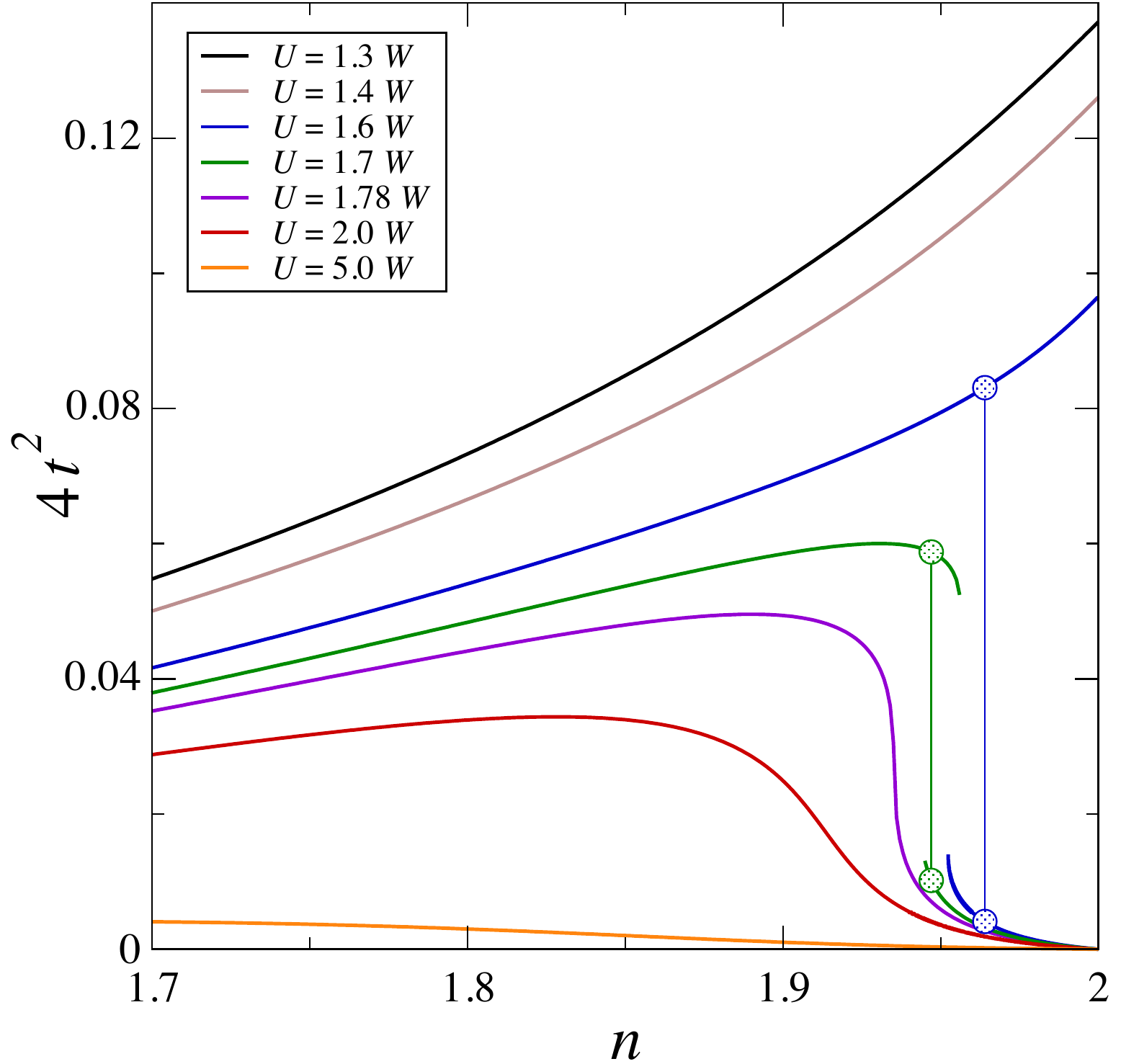}
}\caption{
Single (a) and triple (b) occupancy  in dependence on filling $n$ for $J_{\rm H} = W/6$. The circles and the vertical thin lines characterize the first order transitions.  }
 \label{Fig:p_Jp2}
\end{figure*}

In a fashion similar to the $d$-bosons the critical point $(U_c^*, n^*_{c})$ 
is also central to the density dependence of the $p$ and $t$ bosons; there, they
both exhibit an inflection point with diverging derivative with respect to
$n$. For $ U > U_c^* $ inflection points remain, though less visible, while
the amplitude of the derivatives diminishes when moving away from the critical
point. For $ U < U_c^* $ all bosons jump
at the first order transition, and  a smooth behavior is restored for $ U <
U_c(2)$. 
Similar comments apply to the bosons $p$ and $t$ involving single and triple
occupancy, respectively: as shown in Figs.~\ref{Fig:p_Jp2}(a) and (b) they vanish in
the Mott insulating phase, and exhibit related inflection points at the
critical point $(U_c^*, n^*_{c})$. It should also be noticed that $t$
increases when slightly hole-doping the Mott insulator ($\delta \lesssim
10\%$). This leads to a sizeable gain of kinetic energy for electrons moving
in a background of essentially doubly occupied sites and reinforces the
coherence of the quasiparticles that is lost in the Mott insulating phase.


\section{BEG-parameters}
\label{app_BEG_parameters}

In Section~\ref{sec:BEG} we have set up the BEG scheme as  a phenomenological approach to model the bosonic degrees of freedom. 
In principal, one can devise a microscopic approach through a strong coupling expansion or Schrieffer-Wolff transformation in order to project onto a subspace 
with no triple and quadruple occupied sites. Such an expansion generates nearest-neighbor exchange terms. If it is assumed that only configurations with $d_{\rm P}$ and $d_{\rm A}$ 
are relevant, then the Hamiltonian contribution for double occupancies reads
\begin{align}
{\cal H}_{\rm P,A}=\; &(U -3 J_{\rm H})\sum_{i=1}^{N_{\rm L}} n_{{\rm P}_i} + (U -2 J_{\rm H})\sum_{i=1}^{N_{\rm L}} n_{{\rm A}_i}\notag\\
&-V_{\rm PA}\sum_{\langle i,j \rangle} (n_{{\rm P}_i}n_{{\rm A}_j} +n_{{\rm A}_i}n_{{\rm P}_j})\notag\\
&-V_{\rm PP}\sum_{\langle i,j \rangle} n_{{\rm P}_i}n_{{\rm P}_j} -V_{\rm AA}\sum_{\langle i,j \rangle} n_{{\rm A}_i}n_{{\rm A}_j} \notag\\
&-\mu \sum_{i=1}^{N_{\rm L}} (n_{{\rm P}_i}+n_{{\rm A}_i}) -\mu N_{\rm L}
\label{eq:PA-hamiltonian}
\end{align}
where $n_{{\rm P}_i} $ ($n_{{\rm A}_i}$)  is the number operator for a $d_{\rm P}$-  ($d_{\rm A}$)-configuration on site $i$.
In the expression with the chemical potential one expects $-\mu \sum_{i=1}^{N_{\rm L}} (n_{{ p}_i} + 2n_{{\rm P}_i}+2 n_{{\rm A}_i})$ where $n_{{p}_i}$ denotes the
number operator for singly occupied sites and  factors of two take into account that double occupied sites contribute two electrons. Yet, with the
relation $ \sum_{i=1}n_{{ p}_i} = N_{\rm L} - \sum_{i=1} (n_{{\rm P}_i}+ n_{{\rm A}_i})$ one confirms the last line of Eq.~(\ref{eq:PA-hamiltonian}). 

In Sec.~\ref{sec:BEG} we included the  $d_{\rm H}$-configurations through the constraint
Eq.~(\ref{eq:relds}) however those sites are not represented by a proper term
in the Hamiltonian. This approach is justified if  the number of such sites,
that is $d_{\rm H}^2 $, is much smaller than $d_{\rm P}^2 $, $d_{\rm A}^2 $
and doping $\delta$ which is true close to the considered transition (compare
Fig.~\ref{Fig:ds}(a), (b), (c), and Fig.~\ref{Fig:BosOcc}). If one introduces the on-site energy $U\,\sum_{i=1} n_{{\rm H}_i} $ for the sites with $d_{\rm H}$-configuration then one can identify a shift of $\Delta$, a quantity which is determined below, however this does not affect our results in a
qualitatitive way. 

\begin{figure}[b]
\includegraphics[scale=0.28]{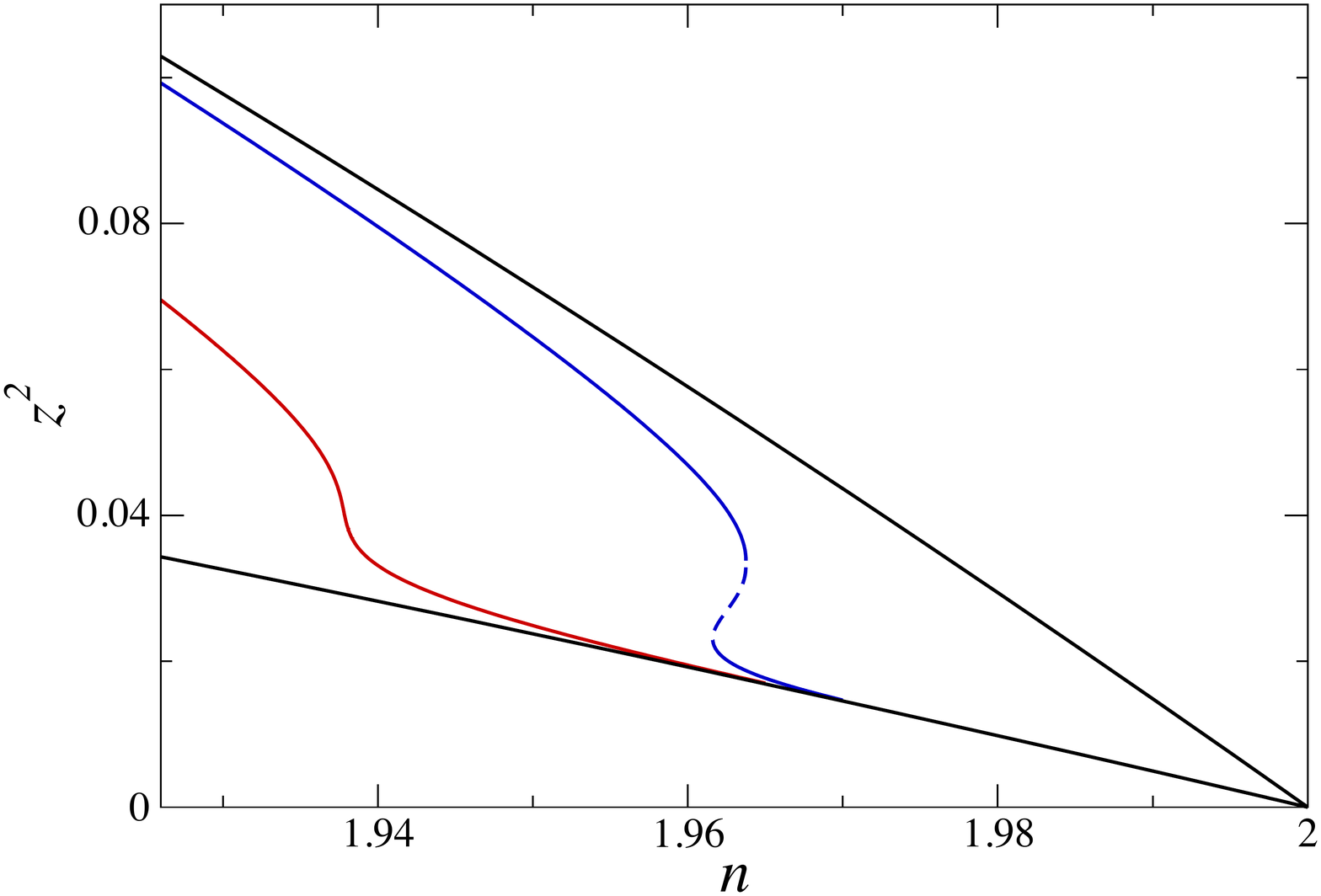}
\caption{Quasiparticle residue from BEG evaluation for finite ${\cal L}$. The black lines are the upper and lower bounds for $z^2$ from Eq.~(\ref{eq:z2-boundaries}). Parameters are   ${\cal J}\! = \!0.0$, ${\cal L}\! = \!-1.0$, and  $g_{\rm fb}\! = \!20$ (red) for the continuous transition, and  ${\cal J}\! = \! 0.1$,  ${\cal L}\! = \!-0.9$, and  $g_{\rm fb}\! = \!180$ (blue) for the discontinuous transition. All energies are in units of $J_{\rm H}/2$ and $h\! = \!1.0$.}
\label{Fig:BEG_z2_l_neq_0}
\end{figure}

Here we do not intend to determine the exchange coupling  parameters $V_{\rm PP}$, $V_{\rm AA}$, and $V_{\rm PA}$ as functions of $U$, $J_{\rm H}$, and $t$ and $t'$ explicitly. We rather discuss qualitatively their dependencies and use them as phenomenological parameters. It is our intention to gain an approximate understanding of the phase transitions identified in slave boson theory within a much  simpler framework. For this purpose we now relate the states of Hamiltonian (\ref{eq:PA-hamiltonian}) to those of the generalized BEG-model (\ref{BEG-H})
through the identification of 
$S_i=1$ with the spin-parallel occupation of the two orbitals $n_{{\rm P}_i}$,  the pseudospin $S_i=-1$ with the spin-antiparallel occupation of the two orbitals $n_{{\rm A}_i}$, and $S_i=0$ with 
the single occupation $n_{{ p}_i}$. This comparison of matrix elements of Hamiltonians (\ref{eq:PA-hamiltonian}) and (\ref{BEG-H}) yields the following relations:
\begin{align}
h= J_{\rm H}/2\,, \quad \Delta=& \,U-\frac{5}{2} J_{\rm H} -\mu\notag\\
{\cal L}=\frac{1}{4}(V_{\rm PP}-V_{\rm AA})\, , \quad {\cal K}=&\frac{1}{4}(V_{\rm PP}+V_{\rm AA}+2\,V_{\rm PA})\notag\\
{\cal J}=\frac{1}{4}(V_{\rm PP}+&V_{\rm AA}-2\,V_{\rm PA})\label{eq:parameters}
\end{align}
Obviously, the magnetic field $h$ of the pseudospin is set by Hund's  coupling $J_{\rm H}$ and the chemical potential related to the pseudospin particles, $-\Delta$, is determined by the chemical potential $\mu$. However, we do not calculate  $\Delta$ through $\mu$ directly but we gain $\Delta$ from the mean-field equation (\ref{eq:MF-delta}).

All exchange coupling parameters $V_{\rm PP/AA/PA}$ are expected to be positive and of order $2 t^2/U$ in the strong coupling regime where not only $t$ but also $J_{\rm H}$ is sizably smaller  than $U$; we also take $(t'/t)^2\ll 1$. 

It is then reasonable to assume that ${\cal L}$ and ${\cal J}$  are considerably smaller than ${\cal K}$ because terms of order $ t^2/U$ cancel in
${\cal L}$ and ${\cal J}$ on account of the minus signs in their respective relations (\ref{eq:parameters}). The coupling ${\cal L}$ is expected to be negative on account of $V_{\rm PP}<V_{\rm AA}$ which results from equal energies of excited states in both, $d_{\rm P}-d_{\rm P}$ and $d_{\rm A}-d_{\rm A}$, and lower energy in the ground state of the $d_{\rm P}-d_{\rm P}$ configuration (so the denominator in the strong coupling expression of the exchange energy is larger for $d_{\rm P}-d_{\rm P}$ than for $d_{\rm A}-d_{\rm A}$). Therefore we conclude that ${\cal L}$ is negative, ${\cal J}$ can have both signs, and ${\cal K}$ is positive and much larger than the absolute value of either ${\cal J}$ or ${\cal L}$.

In Sec.~\ref{subsec:BEGresults} we chose ${\cal L}=0.0$. In Fig.~\ref{Fig:BEG_z2_l_neq_0} we show that a finite negative ${\cal L}$ can produce similar results if the further parameters are chosen properly. In fact, a negative ${\cal L}$ has a similar effect on the orbital magnetization $m$ as a negative ${\cal J}$ if $q$ is close to 1.


\section{BEG-phase-diagram}
\label{app_BEG_phase_diagram}

\begin{figure}[!t]
\includegraphics[scale=0.3]{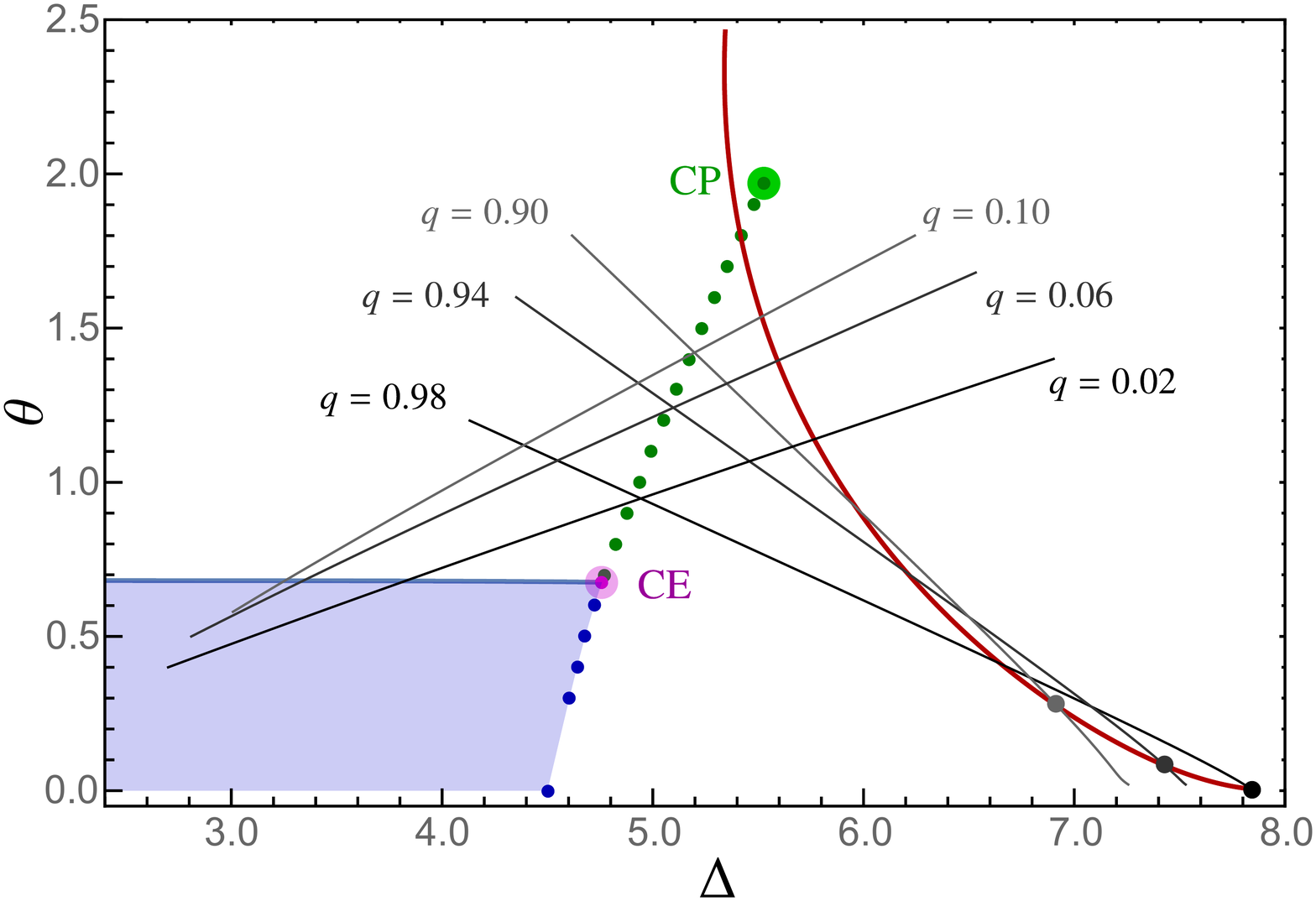}
\caption{BEG phase diagram. The parameters are  ${\cal J}\! = \!-1.0$, ${\cal L}\! = \!0.0$, and ${\cal K}\! = \!8.0$;  these energy parameters as well as temperature $\theta$  and chemical potential $\Delta$ of the pseudospins are in units of the field $h$. The blue line is the continuous phase transition from the antiferromagnetic state at low $\theta$ (blue area) to the paramagnetic state at high $\theta$. The blue dots represent a first order transition line from the antiferromagnetic state to a paramagnetic state with lower values of $q$, and the green dots depict the first order transition line in the paramagnetic state, also from a higher to a lower value of $q$. 
CP is a critical point and CE is the critical end point. The gray curves connect states with fixed $q$-values. The red line is the curve in the parameter space ($\Delta,\theta$) 
for $g_{\rm fb}\! = \!50$
along which we move when we go from large doping ($n=1.8$, that is $q=0.70$ at the upper end point) to small doping  ($n=1.98$, that is $q=0.98$ at the lower end point). 
For this curve $\theta$ is  the effective temperature $k_{\rm B} T_{\rm eff}$. The black dots on that curve are placed at the actual crossing points with the lines of fixed $q$. 
}
\label{Fig:BEG_pd_Delta}
\end{figure}
In Section~\ref{sec:BEG} we considered an antiferromagnetic version of the BEG-model as it allowed to address both, continuous and discontinuous transitions.
The question then arises if the system is actually in an antiferromagnetic state or if the parameter regime is such that the state is still paramagnetic which we have assumed in our evaluation. 

The phase diagram of the spin-1 BEG model was studied by mean field evaluations (see, for example, Refs.~\cite{BEG1971,Saito1981,Hovhannisyan17}), in renormalization-group analyses~\cite{Tremblay84,Bakchich92,Antenucci14}, with exact recursion relations on the Bethe lattice~\cite{Erdinc06}, and  Monte Carlo techniques (see, e.g., Ref.~\cite{Tanaka85,Dani14}), not the least because it reveals a plethora of phase transitions including a tricritical point in a certain parameter range.

In Fig.~\ref{Fig:BEG_pd_Delta} the phase diagram is displayed for a set of
parameters which we used in Section~\ref{sec:BEG}. A tricritical point is
absent for this large value of ${\cal K}$. When Eq.~(\ref{eq:T-profile}) is
solved  for $m$  at given $q$ and the resulting $m(q)$ is inserted in
Eq.~(\ref{eq:MF-delta}), one can identify a curve $T_{\rm eff}(\Delta)$
parameterized by $q$. This is the red curve in Fig.~\ref{Fig:BEG_pd_Delta} where $k_{\rm B}
T_{\rm eff}$ is the temperature $\theta$. 
We also plot lines along which $q$ is constant.
It is only the lower crossing point
of the constant-$q$ lines with the $T_{\rm eff}(\Delta)$-curve that represents
a solution of  Eqs.~(\ref{eq:T-profile}) and (\ref{eq:MF-delta}).

\begin{figure*}[t]
\subfloat[]{
 \includegraphics[scale=0.273]{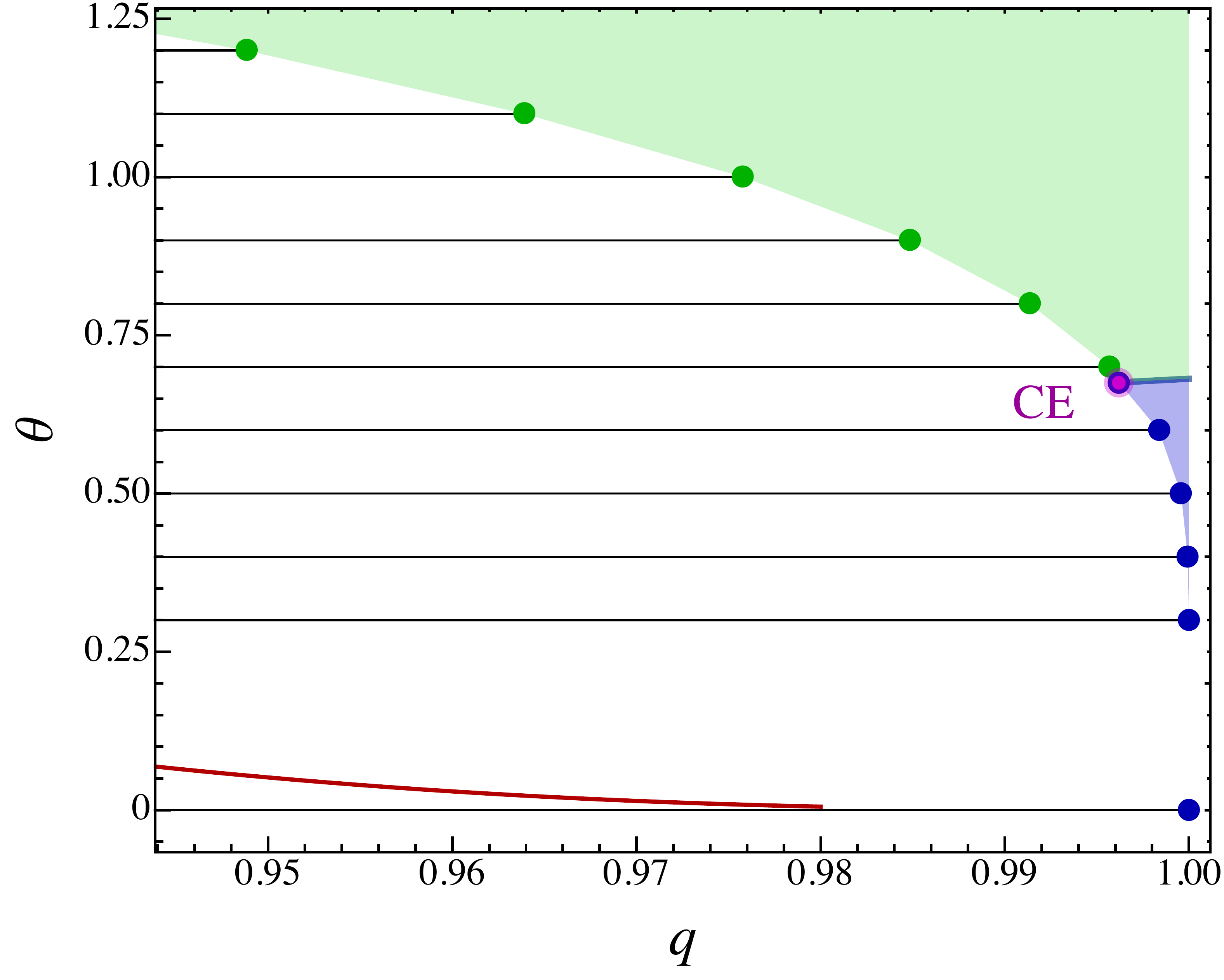}
}\hfil
\subfloat[]{
 \includegraphics[scale=0.265]{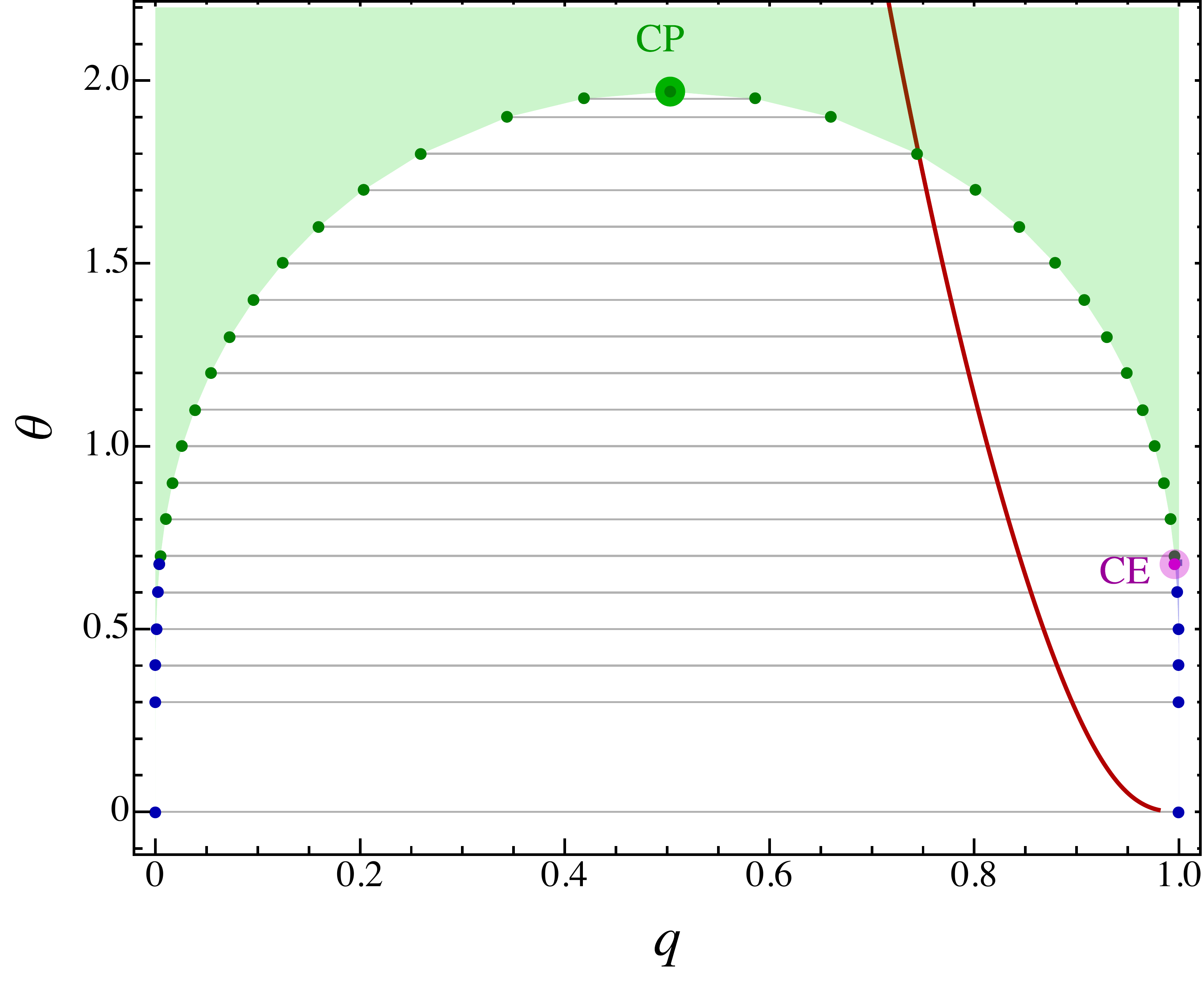}}
\caption{BEG phase diagram in dependence on $q$. The parameters are  ${\cal J}\! = \!-1.0$, ${\cal L}\! = \!0.0$, and ${\cal K}\! = \!8.0$; these energy parameters as well as
the temperature $\theta$  of the pseudospins are in units of $h$. The blue triangular area represents the antiferromagnetic state (see panel (a)). The blue line depicts the continuous transition, the blue dots the  first order transition  from the antiferromagnetic state to a paramagnetic state, and the green dots refer to the first order transition in the paramagnetic state. The gray lines between these dots connect the equilibrium states with high $q$ to those with low $q$ at the first order phase transition (see panel (b)). States within the white area are not characterized by a global minimum of the BEG free energy. The green area is the regime where global minima of the free energy exist.
CP is a critical point and CE is the critical end point.  The red curve is
given by $k_{\rm B} T_{\rm eff}(q)$ for $g_{\rm fb}\! = \!50$. 
}
\label{Fig:BEG_pd_n}
\end{figure*}

Note that the slope of the constant-$q$ lines changes sign when going from $q$ close 1 to small values of $q$. At the first order transition  (green dots) the high-q lines  cross with appropriate low q-lines as there the $q$-value jumps when the thermodynamic equilibrium is considered (for example, a jump from $q=0.98$ to approximately $q=0.02$). In fact, high values of $q$ represent the thermodynamic equilibrium on the left hand side of the transition (that is, for lower $\Delta$) and low values of $q$  are in equilibrium on the right hand side of the transition (that is, for larger $\Delta$).

From these considerations it is obvious that the states that we find as solutions from the mean-field equations are not equilibrium states of the pseudospin system: above approximately $q=0.75$
(for  $g_{\rm fb}\! = \!50$) the red curve is  in the parameter regime where low values of $q$ would be (global) equilibrium solutions whereas we are forced to realize high-$q$ solutions because the filling $n$, which is approximately $1+q$, approaches half-filling at $q=1$ and the effective temperatures goes to zero.

The choice of the parameter space $(\Delta,\theta)$ yields the conventional representation of the phase diagram of the BEG model but here it appears to be more appropriate to consider functions of $q$ instead of $\Delta$ on account of the necessity to specify the filling. Correspondingly we display in Fig.~\ref{Fig:BEG_pd_n} the phase diagram in the parameter space $(q,\theta)$. It can be clearly seen that the $k_{\rm B} T_{\rm eff}(q) $-curve is placed in the nonequilibrium regime (white area) except for high temperature and $q$ below 0.75 (green area). The antiferromagnetic regime (blue area) appears as a small triangle
in this parameter space and it is well separated from the states that are relevant for the analysis of the phase transitions discussed in the present framework.


\begin{references}
\bibitem{comment1} One may expect that the energy scale $T_{\rm eff}$ results from a RG-type evaluation.
\end{references}
\end{document}